\documentclass[11pt]{article}
\usepackage{amssymb}
\usepackage{graphics,graphicx}
\usepackage{color}
\usepackage{amsbsy}
\usepackage[left=2cm,right=2cm,top=2cm,bottom=2cm,bindingoffset=0cm]{geometry}
\usepackage{amsmath}
\usepackage{ccaption}
\usepackage{epstopdf}
\usepackage{verbatim}
\usepackage{dsfont}
\usepackage{soul}
\usepackage{enumerate}

\title{Dynamic fracture of a discrete media under moving load.}

\author{Nikolai Gorbushin, Gennady Mishuris\footnote{corresponding author: ggm@aber.ac.uk}
\\
{\it Department of Mathematics,
Aberystwyth University, }
\\ {\it Ceredigion SY23 3BZ, Wales, UK}
}
\date{}

\begin{document}

\maketitle

\begin{abstract}
Most of the research concerting crack propagation in discrete media is concerned with specific types
of external loading: displacements on the boundaries, or constant energy fluxes or feeding waves originating from infinity. In this paper the action of a moving load is analysed
on the simplest lattice model: a thin strip, where the fault propagating in its middle portion as the result of the moving force acting on the destroyed part of the structure.
We study both analytically and numerically how the load amplitude and its velocity influence the possible solution, and specifically the way the fracture process reaches its steady-state regime. We present the relation between the possible steady-state crack speed and the loading parameters, as well as the energy release rate.
In particular, we show that there  exists a class of loading regime corresponding to each point on the energy-speed diagram (and thus determine the same limiting steady-state regime).
The phenomenon of the "forbidden regimes" is discussed in detail, from both the points of view of force and energy.
For a sufficiently anisotropic structure, we find a stable steady-state propagation corresponding to the "slow" crack.
Numerical simulations reveal various ways by which the process approaches - or fails to approach - the steady-state regime.
The results extend our understanding of fracture processes in discrete structures, and reveal some new questions that should be addressed.

\end{abstract}


\section{Introduction}
Theoretical works on a crack propagation in structured media have revealed various phenomena that are not observable when considering the cracks in an elastic continuum. One of the major observations following from the study of cracks in a lattice is that the static crack becomes unstable by application of displacements which almost twice exceed the size predicted by using the energy criterion; this effect was referred to as as lattice trapping in \cite{thomson1971}. The development of a consistent theory of crack propagation in such structures originates in work by Slepyan \cite{slepyan1984} for the Mode III crack (rectangular lattice) and the Mode I and II \cite{Kul1984} (triangular lattices), leading eventually to a fully comprehensive study in \cite{slepyan2012}. The proposed methods appeared to be extremely efficient in examining various fracture problems and capable of explaining various related phenomena \cite{kessler1999,marder1995,slepyan2012}. In particular, apart from explaining trapping in various lattice structures \cite{colquitt2012,slepyan2012},
it was also instrumental in recognizing the role of the dissipation mechanism in fracture mechanics \cite{liu1991,sharon1996,slepyan2012} in the description of a crack propagation in discrete and structural waveguides \cite{brun2013,carta2014,mishuris2012} and the analysis of the phase transitions and bistable structures \cite{cherkaev2010,trofimov2010,truskinovsky2005,truskinovsky2006}. The method is equally efficient for structures of distinct geometries (rectangular and triangle lattices), fracture modes, for both open cracks and bridge cracks \cite{Mish2007,Mish2008}, and both homogeneous and inhomogeneous structures \cite{Mish2007b,Mish2009,nieves2013}. Although most of the works so far have been concerned with the structures constructed as masses linked by elastic springs, structures where the links are elastic beams have been recently analysed \cite{nieves2016,ryvkin2010}.

Crack propagation instability and fast crack branching has been a long standing problem of fracture considered in the framework of classical elasticity~\cite{mishuris2010dynamic,mishuris2006steady,movchan1998perturbations,willis1997three} and couple stresses~\cite{morini2014remarks,morini2013fracture}. The approach suggested by Slepyan, supplemented by extensive numerical simulations and experimental analysis, has allowed to address this problem from the microscopic point of view~\cite{fineberg1999,marder1995}. Some "forbidden regimes" have also been identified, explaining the instability of crack propagation for low crack speed, while the "admissible regimes" corresponding to possible steady-state crack propagation have been discussed for various (both rectangular and triangular) lattices \cite{behn2015,gorbushin2015,gorbushin2016,marder1995} with moderate and fast speeds. Moreover, for the fast propagating crack, a branching phenomenon appears as a result of possible irregular breakage of the links \cite{marder1995}. Other phenomena recently discovered and explained include clustering and forerunning regimes, as observed in differing lattice structures \cite{ayzenberg2014,gorbushin2016,nieves2016,nieves2017}.
Recently, the lattice structure approach has been used to model the complex phenomenon of hydraulic fracture \cite{marder2015}. Additionally, Slepyan's approach allowed to study the consequences of application of different dynamic fracture conditions, e.g. the incubation time criterion~\cite{gorbushin2016effect}, for the discrete mechanical systems~\cite{gorbushin2017influence}.

Experimental results \cite{ivankovic2004,sharon1999} and numerical simulations on cellular and lattice structures \cite{lipperman2007,marder1995,Mish2009,nieves2016,nieves2017} as well as elastic media \cite{parisi2002}
show that the steady-state regimes predicted by the theory can be reached.
However, the validity of the solution found using the analytical models always should be always verified via both numerical simulations and experimentation as said solution
is always obtained under the assumption of the existence of the steady state regime. A real solution of the problem may be different to that predicted steady state example
(for example, the regular cluster propagation regime discovered numerically \cite{Mish2009} and proved later analytically \cite{slepyan2010} is a simple but illuminating alternative).

In spite of the fact that the aforementioned models describe a variety of fracture events, there are unfortunately open questions that remain unaddressed.
Thus, most of the research to date considers steady state crack propagation in discrete media appearing as the result of the actions of very limited types of external loading: displacements on the boundaries, and constant energy fluxes and feeding waves from infinity.
Varying the choices for the loading parameters can lead to different outcomes. Even for a static problem in a lattice structure loaded by both external and internal forces, a kind of material softening behaviour has been
predicted \cite{mish2014}. It is clear that for dynamic problems, which are essentially non-linear, there are complex behaviour, and that each load configuration should be considered separately.

In the present paper we discuss crack propagation as the result of a an applied force moving with constant speed and amplitude. We analyse the impact of the loading parameters (force magnitude and velocity of the force location) on the fracture process (character of the crack propagation, whether it approaches the steady-state regime predicted by theory, etc.). We demonstrate that, even in the areas considered to be understood, such as the forbidden propagation regimes, our knowledge is incomplete. We show that transient regimes may approach the same steady-state quite differently depending on the combination of loading parameters. We compare the advantages of the {\it energy-speed} and {\it load-speed} approaches, and we discover new stable slow propagation regimes for highly anisotropic structures.

The work revisiting the study of crack propagation in discrete structures~\cite{slepyan2012,slepyan1984} has a fundamental character as it further explains the phenomenon of stable crack propagation in such materials in conjecture with the way how the load is applied to the system. A direct relationship between the load and the developed crack speed has been established. A straightforward extension to more complex structures and loading conditions can be performed. The proposed model, even in its simplest formulation, allows to study the sensitivity of steady state crack movement on the loading parameters. Bearing in mind a wide range of applications for various discrete structures (starting from the classic bridge constructions~\cite{brun2013} to modern matamaterials~\cite{craster2012}), analysis of possible catastrophic events (dynamic fracturing/decomposition of the structures) is a crucial task to guarantee safety of modern constructions and new materials. Moreover, such discrete model could be also useful for modelling of surface phenomena localized near the crack such as surface elasticity models with structured coatings~\cite{eremeyev2016effective}. Let us note that the problems of dynamic response under moving loads and masses are widely analysed in the literature, see ~\cite{fryba2013,gavrilov2016revisitation}, for instance. Thus, the proposed model can enrich the study in this field and adapted for the other configurations.

The structure of the paper is as follows. We start with a very detailed numerical simulation of a selected geometrical, mechanical and loading configuration of the problem. This
allows us to highlight the main peculiarities of the fracture process as well as to indicate the main challenges for the numerical simulations.
Here, we also explain in great detail how we evaluate the predicted steady-state speed of the crack from the numerical analysis and discuss the accuracy of the method.
Then we revisit the theoretical analysis of the problem, completing it for the case of a moving force with constant speed.
We present only the main steps of the analysis required to understand the new features of the method.
Some purely technical derivations are moved into the Supplementary material. Finally, we examine those results for more intensive numerical simulations, confirming the main finding and revealing/posing new questions.

\section{Preliminary numerical simulations.}
\label{section:NumericalSettings}
\subsection{Numerical set up.}
\begin{figure}[h!]
\center{\includegraphics[scale=0.8]{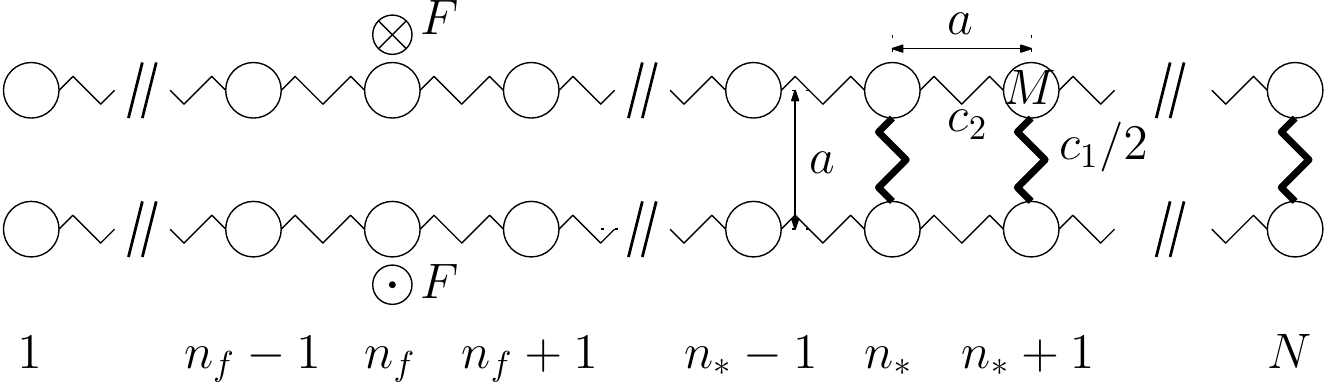}}
\captiondelim{. }
\caption[ ]{Chain of oscillators with equal masses $M$ connected together by linear springs of stiffness $c_2$ (normal lines) and to the rigid foundation with springs of stiffness $c_1/2$ (fat lines). The crack position is defined by an oscillator with index $n_*$. The force $F$ is applied out of plane to the upper row and in to the plane on the lower oscillator, where $n_f$ is its position. The vertical springs of stiffness $c_1/2$ consequently break as the crack moves. $a$ is an equilibrium distance between the oscillators.}
\label{fig:Infinite Chain}
\end{figure}

Let us consider the propagation of a crack between two symmetrical rows of oscillators under anti-plane deformation (mode III fracture).
The length of the structure is defined by the number $N$ of oscillators in one row. Forces of constant magnitude $F$ are applied outwards and inwards to the top and bottom rows, respectively, for identically situated forces. The described configuration is shown in Fig.\ref{fig:Infinite Chain} where all the mechanical quantities (masses, spring stiffnesses) are also shown.

Utilising the symmetry of the problem under consideration, i.e. the applied load and physical parameters, the linearised equations of motion of such a system take the form:
\begin{equation}
\begin{gathered}
M\ddot{u}_n(t)=c_2(u_{n+1}(t)+u_{n-1}(t)-2u_{n}(t))+F\delta_{nn_f},\quad 1<n<n^*,\\
M\ddot{u}_n(t)=c_2(u_{n+1}(t)+u_{n-1}(t)-2u_{n}(t))-c_1u_{n}(t),\quad n_* \leq n<N,
\end{gathered}
\label{eq:ProblemNumerics}
\end{equation}
where $M$ is the mass of an oscillator, $c_1$ the double stiffness of the springs that break while the crack propagates, $c_2$ is the spring constant of the links between neighbouring oscillators, $F$ is the magnitude of an external force, $n_*=n_*(t)$ is the position of the crack tip, $n_f$ is the location of the applied force, $u_n(t)$ is the outward displacement of an oscillator with index $n$ of the top row. The discrete Kronecker delta is written as $\delta_{nm}$. The displacements of oscillators from the top and bottom rows with the same index have the same magnitudes but different signs.

The initial conditions for the problem are set to be homogeneous:
\begin{equation}
u_n(t)=0,\quad \dot{u}_n(t)=0,\quad t=0,\quad 1\le n\le N.
\label{eq:ICSNumerics}
\end{equation}

The numerical simulations also require boundary conditions to be stated as well, such as clamped or free ends of the chains. However, we are interested in the analysis of a solution close to a crack tip. for the choice of a reasonably large number of $N$ oscillators the displacements of oscillators close to a crack tip do not depend on the stated boundary conditions. We will later explicitly show the effects of boundary conditions in one particular case.

In the presented configuration, we assume that the crack propagates from the left to the right. The displacement at the crack tip is subjected to a deformation fracture criterion given in the following form:
\begin{equation}
\begin{gathered}
u_{n_*}(t_*)=u_c,\\
u_n(t)<u_c,\quad n>n_*(t),
\end{gathered}
\label{eq:FractureCondition}
\end{equation}
where $u_c$ is a constant and the second condition is consistent with the assumption that the crack tip can be uniquely defined by index $n_*$.

The model also provides a parameter for the critical value of the crack speed $v$ which is defined by the value of a speed of sound $v_c$ of the broken part of the structure:
\begin{equation}
v<v_c=\sqrt{\frac{c_2}{M}}a.
\label{eq:CriticalSpeed}
\end{equation}

This limitation follows from the evidence that the load is applied far away behind a crack tip (from the left to it as shown in Fig.\ref{fig:Infinite Chain}) and has to continuously provide the energy supply for the crack propagation. It remains valid as long as the force remains situated in the broken part of a chain. In the case where the load moves faster then the crack tip, this condition is guaranteed by the computational time frame.

We allow the location of the force $n_f$ to vary according to the following rule:
\begin{equation}
n_f(t)=n_f(0)+\frac{v_f}{a}t,\quad v_f=const.
\label{eq:Position_of_Force}
\end{equation}
For the computations we need to have integer values for $n_f(t)$, and thus choose the ceiling of this number. We also trialled using the floor of $n_f(t)$ or its more general rounding, but the change did not seem crucial in the prediction of the steady-state crack speed. In further analysis, $v_f=0$ corresponds to a fixed load, $v_f>0$ indicates that the force is moving toward the crack tip, $v_f<0$ that the force is moving in the opposite direction.
In the following analysis we normalise the velocities by the equilibrium distance between the oscillators:
\begin{equation}
\tilde{v}=\frac{v}{a},\quad \tilde{v}_c=\frac{v_c}{a},\quad \tilde{v}_f=\frac{v_f}{a}
\label{eq:Normalisation}
\end{equation}
From now on we use the normalised variables shown in \eqref{eq:Normalisation} omitting the tildes for practicality. The numerical simulation is performed by running series of iterations. Let us define $t_*^j$ as the time of the $j$-th fracture event at the point $n=n_*^i(t_*^j)$. The $j$-th iteration then has the following steps:
\begin{enumerate}
\item Once condition \eqref{eq:FractureCondition}$_1$ is fulfilled, the relevant solution for moment $t_*^j$ is archived for all values of $n$:
\begin{equation}
u_n^j=u_n(t_*^j),\quad w_n^j=\dot{u}_n(t_*^j).
\label{eq:NewICS}
\end{equation}

\item The spring of stiffness $c_1/2$ between the oscillators with index $n_*^{j}(t_*^j)$ is removed, and for the next iterations we choose $n_*^{j+1}(t)=n_*^{j}(t_*^j)+1$. At this point we also check condition \eqref{eq:FractureCondition}$_2$, but in the cases considered it was always already fulfilled.

\item System \eqref{eq:ProblemNumerics} is solved again, using the previously stored values $u_n^j$, $w_n^j$ from \eqref{eq:NewICS} as initial conditions.
\end{enumerate}

All computations are done within the Matlab R2015b environment. The geometrical settings of the structure used in the computations in this section are summarised in Table \ref{table:TableSettings}.
\begin{table}[!ht]
\vspace{2mm}
\begin{center}
\begin{tabular}{|l|c|c|c|c|}
\hline
& $\mathbb{S}_1$ &$\mathbb{S}_2$ &$\mathbb{S}_3$ &$\mathbb{S}_4$\\ \hline
Total number of oscillators, $N$ & 4000 & 4000 & 4000 & 8000\\ \hline
Total number of breakages, $I$ & 1000 & 1000 & 1000 & 1000\\ \hline
Initial crack tip position, $n_*(0)$ & 2000 & 2000 & 2000 & 2000\\ \hline
Initial force position, $n_f(0)$ & 1000 & 1500 & 1900 & 1000\\
\hline
\end{tabular}
\end{center}
\captiondelim{. }
\vspace{-3mm}
\caption[ ]{Geometrical settings of the structure on Fig.~\ref{fig:Infinite Chain} used in the computations.}
\label{table:TableSettings}
\end{table}
The chosen sets of the parameters, $\mathbb{S}_j$, guarantee that the fracture process exhibits stable and well developed behaviour for a sufficiently long time, thus allowing us to study its properties. Comparing results for sets $\mathbb{S}_1$ (shorter structure) and $\mathbb{S}_4$ (longer structure) allows us make some conclusions on the influence of the distance between the initial crack tip (and the point where the force is applied) and the left hand side of the structure. We check the impact of the initial force position $n_f(0)$ on the results with respect to configurations $\mathbb{S}_1$, $\mathbb{S}_2$ and $\mathbb{S}_3$, where the distance decreases with each respective set. We do not employ damping in the numerical computations, but control the overall time in the process before the fracture is affected by the reflected waves approaching the crack tip from the left-hand side and right-hand side of the structure. As mentioned earlier, we will also investigate the influence of the boundary conditions at the ends of the structure.

The following choices for physical parameters remain unchanged throughout all the simulations: $$c_2=1 [F/L], M=1 [m], a=1 [L], u_c=1 [L]$$.
Henceforth, we omit the units in given quantities, assuming them to be appropriate in form.

\subsection{Computation of the crack speed}

In this section we describe the data analysis used throughout the paper to extract the physical and geometrical properties of the process (crack speed, displacement profiles, etc.)
This analysis provides enough confidence to allow us to make conclusions and explain the basic peculiarities of the process. In particular, it establishes a proven link between the results obtained numerically from the discrete structure and those evaluated analytically from the corresponding continuous structure.

One of the most important parameters describing the fracture process is the instantaneous speed of the propagating crack, $v(t)$, which takes discrete values since the structure itself is discrete. Assuming that the crack tip moves by breaking of preceding springs only, i.e. without any breakage being detected ahead, we define an instantaneous crack speed (normalised by the equilibrium length $a$ as in \eqref{eq:Normalisation}) in the following way:
\begin{equation}
v(t_*^j)=\frac{n_*(t_*^{j+1})-n_*(t_*^{j})}{t_*^{j+1}-t_*^{j}}.
\label{eq:CrackSpeed}
\end{equation}
Here $j$ is the number of the latest breakage in the fracture process.

In order to compare the analytical result for the steady-state speed, $v$, which is a constant value for the given geometrical and physical parameters, with the results of the numerical simulations, we need to have an equivalent definition for this quantity, supplemented by a quantitative estimate of the latter.

Although the distribution of the data is not necessarily normal,
we may accept the mean value, $\bar{v}$, of the instantaneous speed, $v(t_*)$, as a possible numerical definition of the limiting steady state crack speed, $v$.   With this in mind, we consider the set of the data starting from the $m$-th breakage of the link with index $j=n_*(0)+m$ , where the remaining part of the fracture process is computed up to the final point $j=n_*(0)+I$, and the instantaneous speed $v(t_*)$ demonstrates a regular oscillatory behaviour with a small amplitude.
\begin{equation}
\bar{v}=\frac{1}{I-m}\sum_{j=m}^{I-1}v(t_*^j).
\label{eq:ComputedSpeed}
\end{equation}

We also may calculate the sample standard deviation $\sigma(v)$, to have some quantitative measure providing an insight into the accuracy of the chosen assumption:
\begin{equation}
\sigma(v)=\sqrt{\frac{1}{I-m-1}\sum_{j=m}^{I-1}\Big(v(t_*^j)-\bar v\Big)^2}.
\label{eq:ComputedSpeed_2}
\end{equation}
An alternative method for estimating the crack speed from the numerical analysis would be to use the average speed on the same interval:
\begin{equation}
v_a=\frac{n_*(t_*^{I})-n_*(t_*^{m})}{t_*^{I}-t_*^{m}},
\label{eq:ComputedSpeed_1}
\end{equation}
where the difference between the values of $\bar v$ and $v_a$ serves as an additional accuracy measure.

\begin{figure}[h!]
\minipage{0.5\textwidth}
\center{\includegraphics[width=\linewidth] {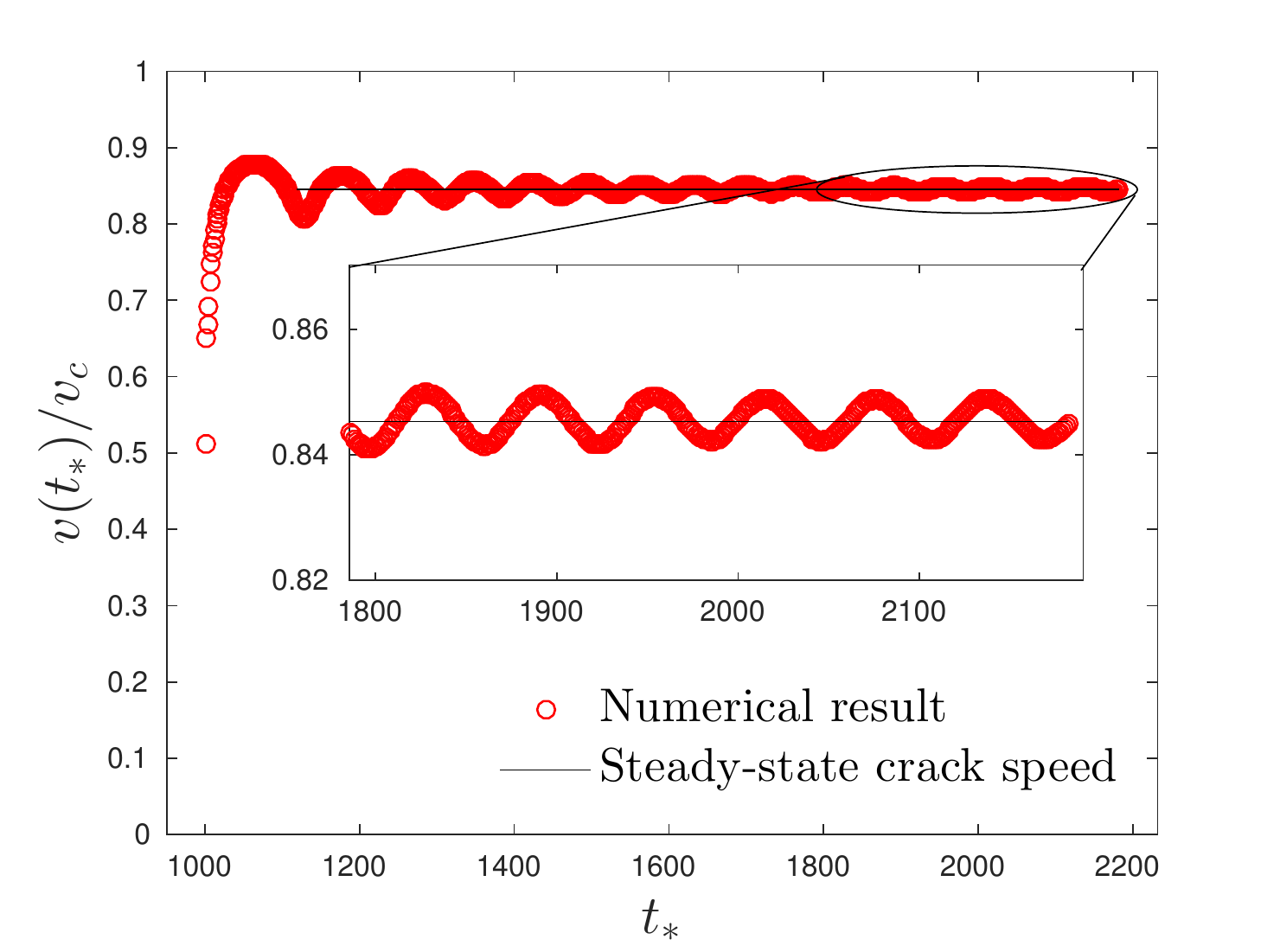} \\ a)}
\endminipage
\hfill
\minipage{0.5\textwidth}
\center{\includegraphics[width=\linewidth]{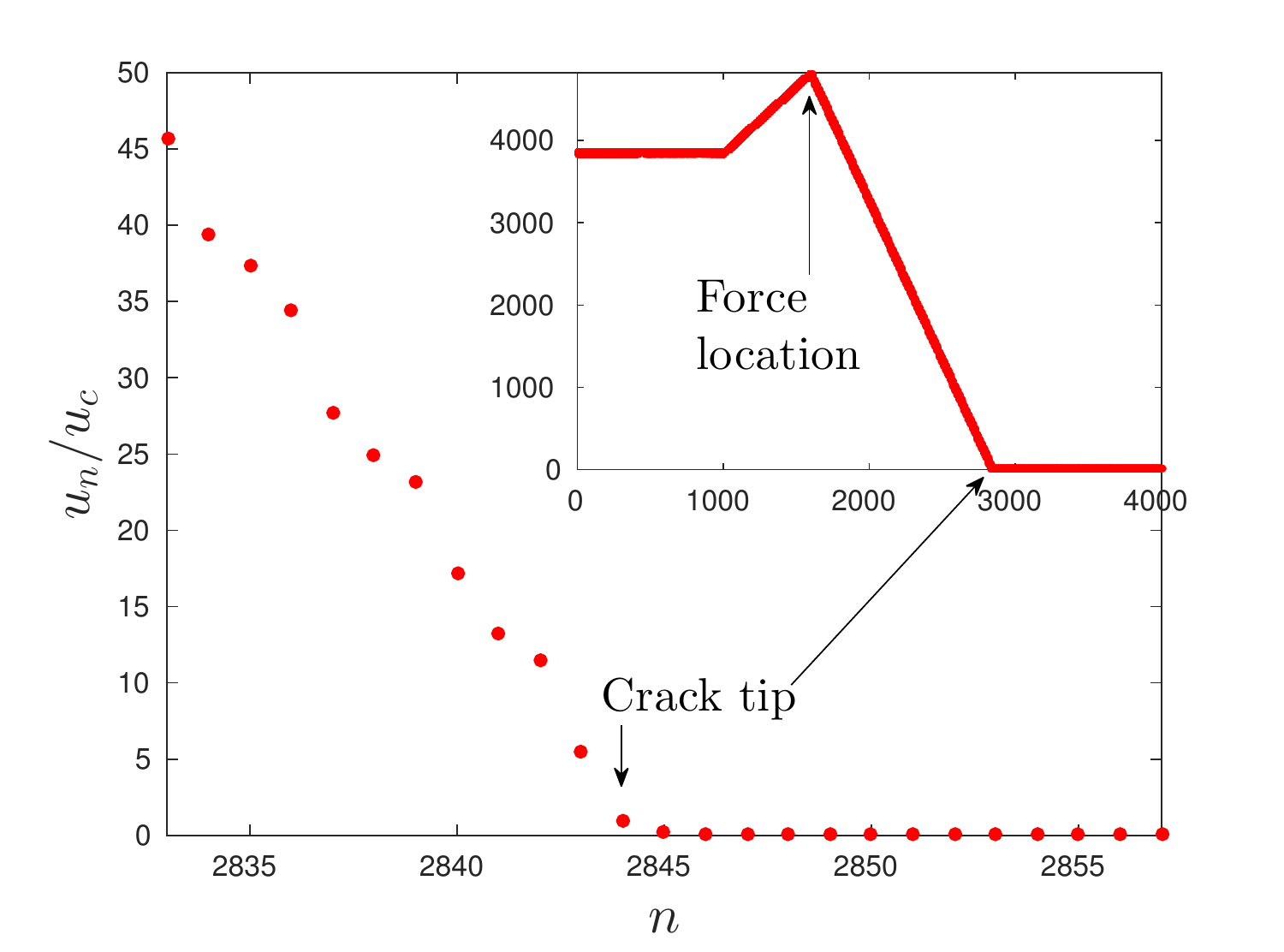} \\ b)}
\endminipage
\captiondelim{. }
\caption[ ]{Results of the computations for geometrical setting $\mathbb{S}_1$ of Table~\ref{table:TableSettings}, where $c_1=2c_2$, $F=5Mu_cv_c^2$ and $v_f=0.3v_c$. Free edge boundary conditions are prescribed at the structure ends: Fig.\ref{fig:example}a) -- The instantaneous crack speed $v(t_*)/v_c$ given by \eqref{eq:CrackSpeed}.
The insert highlights the final stage of the computations, Fig.\ref{fig:example}b) -- The displacements profile of the oscillators close to the crack tip at time $t_*\approx2000$, taken from the middle of the region shown in the insert, during the well established regime shown in in Fig.\ref{fig:example}a).}
\label{fig:example}
\end{figure}

We now analyse the consequences of particular choices for the geometrical parameters when computing the crack speed from the numerical data.
A typical plot for the instantaneous speed, $v(t_*)$, can be seen in Fig.~\ref{fig:example}a), where the typical sample set of the data is one where the oscillations of $v(t_*)$ become regular. This set of data is later
used for evaluation of the steady-state crack speed from the numerical data.

The geometrical configuration used in this example corresponds to set $\mathbb{S}_1$ from Table~\ref{table:TableSettings}, where $c_1=2c_2$, $F=5Mu_cv_c^2$ and $v_f=0.3v_c$.
At both ends of the structure, free boundary conditions are
prescribed. It is clear that the instantaneous speed is not a constant but has a clear tendency to approach some limiting value with time as the fracture process develops.

The profile of the entire structure at a certain moment of the fracture event is shown in Fig.\ref{fig:example}b).
We can observe that the displacements behind the crack tip do not form the pure inclined straight line that is seen when examining the global picture in the insert of Fig.\ref{fig:example}b). This discrepancy is caused by the reflection of waves from the crack tip back to the source. It can be also seen that the amplitude of these waves is much larger than those transmitted into the structure on the crack line ahead (if those exist at all, which is not obvious on the presented scale).

Different strategies can be employed to numerically evaluate the steady-state crack propagation speed from the computations. In Table~\ref{table:TableData_speed}
we present results obtained from three sets of samples
(differing by length of the observation time or the length of the fractured structure) for the same structure $\mathbb{S}_1$.
The shortest period ($m=300$) seems the most appropriate choice when analysing computations done in accordance with equations \eqref{eq:ComputedSpeed} -- \eqref{eq:ComputedSpeed_1}, but it is difficult to make a stronger justification. To illustrate this point,
the speed of the steady-state propagation computed via the analytical formula derived in the next section is $v=0.8457v_c$ (compare with the values in Table~\ref{table:TableData_speed}).
For a reason which will become clear later we will use the largest data set ($m=100$), that contains practically the entire fracture regime except its initial stage.
While sacrificing a little accuracy in the steady-state speed evaluation
we can guarantee in this way not to miss any essential features of the process when the oscillatory behaviour changes (for other sets of the material parameters).

\begin{table}[!ht]
\vspace{2mm}
\begin{center}
\begin{tabular}{|c|c|c|c|c|}
\hline
Starting point & Sample length&$\bar{v}/v_c$ &$v_a/v_c$ &$\sigma(v)/v_c$ \\ \hline
$m=100$ & $I-m=900$ & 0.8452 &0.8453&0.0079\\ \hline
$m=200$ & $I-m=800$&0.8458 &0.8459&0.0051\\ \hline
$m=300$ & $I-m=700$ &0.8456&0.8456&0.0042\\
\hline
\end{tabular}
\end{center}
\vspace{-3mm}
\captiondelim{. }
\caption[ ]{Evaluation of the predicted steady-state crack speed using the formulae \eqref{eq:ComputedSpeed} -- \eqref{eq:ComputedSpeed_1}
and the standard deviation of this value, $\sigma(v),$ for the data presented
in Fig.~\ref{fig:example}a).}
\label{table:TableData_speed}
\end{table}

Another direct conclusion from this preliminary analysis is that the difference between the mean value, $\bar{v}$, and the average value, $v_a$, of the crack speed is definitely
smaller than the accuracy of the computations, bearing in mind its sensitivity with respect to the choice of sample set.
For this reason, from now on we report only the mean values, $\bar{v}$, defined numerically by \eqref{eq:ComputedSpeed}.

In the next subsection we discuss the effects of the choice of geometrical configuration from Table \ref{table:TableSettings} and its impact on the evaluation of the major process parameters.

\subsection{Effect of values of the geometrical and physical parameters}

\begin{figure}[h!]
\center{\includegraphics[scale=0.6]{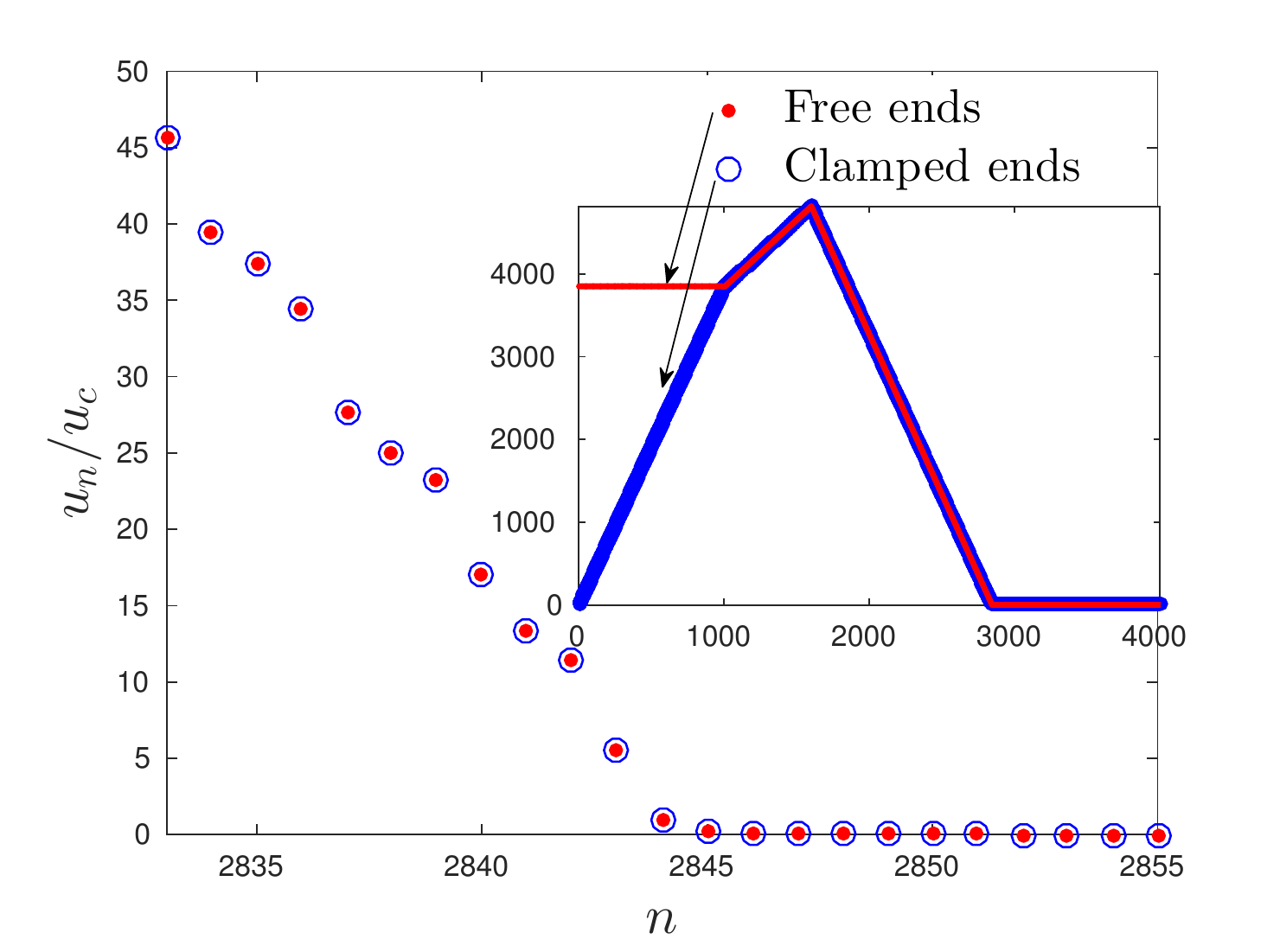}}
\captiondelim{. }
\caption[ ]{Displacement of the oscillators close to a crack tip for two different boundary conditions, given the configuration $\mathbb{S}_1$ from Table\ref{table:TableSettings} and $F=5Mu_cv_c^2$, $v_f=0.3v_c$ and $c_1=2c_2$ at time $t_*\approx2000$.
The insert shows the displacement of the whole chain. The red colour corresponds to free ends, while blue corresponds to clamped ends.}
\label{fig:DifferentBoundaryConditions}
\end{figure}

Firstly, we analyse the impact of the prescribed boundary conditions at the ends of the structure on the numerical results. We consider two options:
free ends and clamped end conditions. The results for the displacement field close to a crack tip and for the entire structure are shown in
Fig.\ref{fig:DifferentBoundaryConditions} for the same geometrical setting, $\mathbb{S}_1$, and  $F=5Mu_cv_c^2$, $v_f=0.3v_c$ and $c_1=2c_2$ at time $t_*\approx2000$.
We observe that, for the chosen numbers of oscillators and iterations, the boundary conditions do not have an effect on the results for the displacement field close to a crack tip, nor those for the crack speed.
The predictions for the inclination slope behind the crack tip are also not affected.

\begin{figure}[h!]
\center{\includegraphics[scale=0.6]{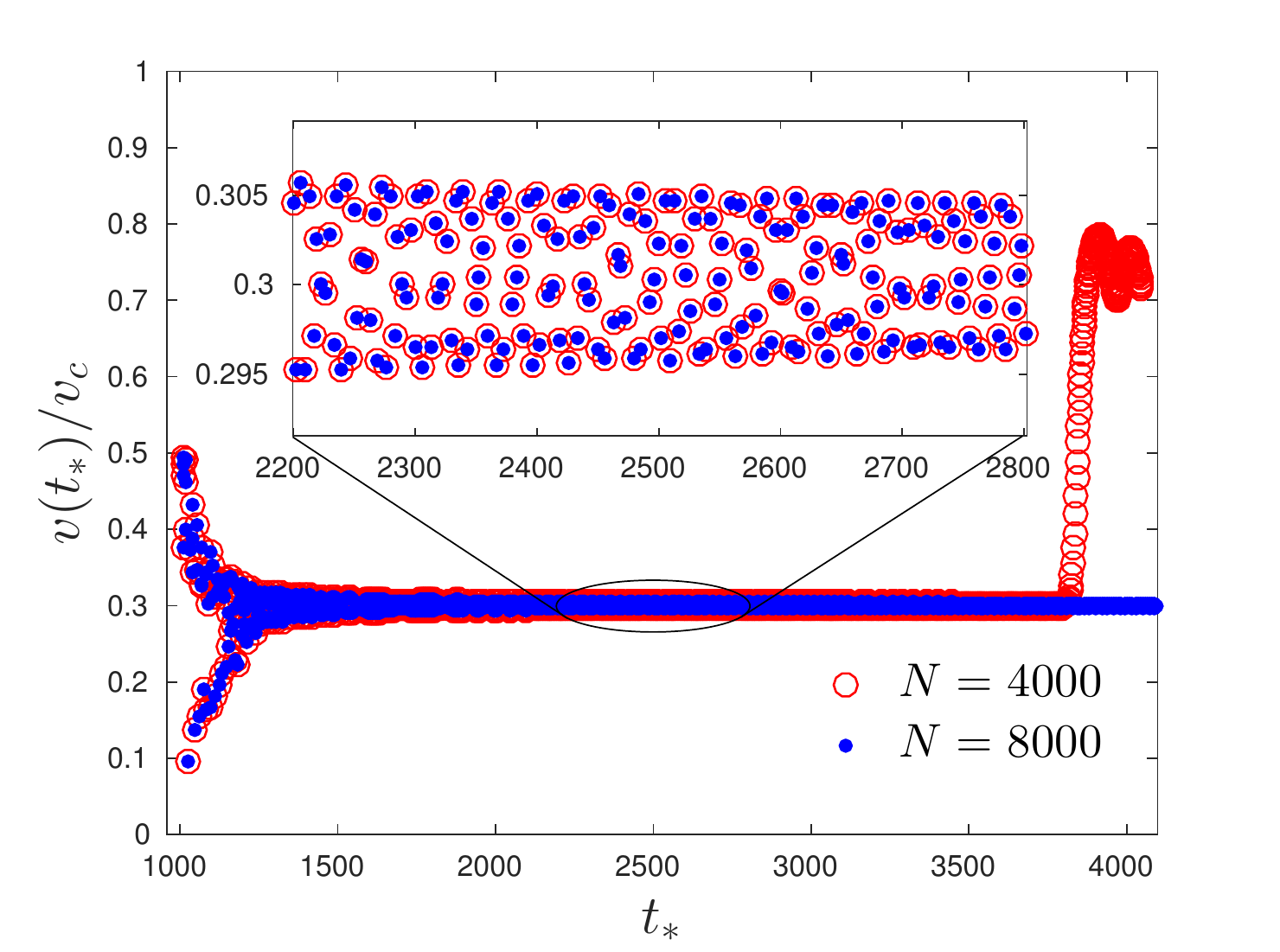}}
\captiondelim{. }
\caption[ ]{The Instantaneous crack speed $v(t_*)/v_c$, given by \eqref{eq:CrackSpeed}, for different total number of oscillators $N$, where $F=1Mu_cv_c^2$, $v_f=0$ and $c_1=0.5c_2$. The results for the set of geometrical parameters shown in red correspond to structure $\mathbb{S}_1$, while in blue  correspond to structure $\mathbb{S}_4$.}
\label{fig:DifferentN}
\end{figure}

The response to the boundary conditions may, however, be noticed if the crack speed is sufficiently slow and the reflected wave reaches the crack tip in the chosen time frame ($I=1000$ fracture events).
This can be avoided by an increase in the number of oscillators in the structure, for example from the number in set $\mathbb{S}_1$ to that in set $\mathbb{S}_4$, both given in Table~\ref{table:TableSettings}. In order to demonstrate this effect, we choose the alternative values for the material parameters: $c_1=0.5c_2$, $F=5Mu_cv_c^2$ and $v_f=0$, leading to a lower steady-state crack speed.  The ensuing results are shown in Fig.\ref{fig:DifferentN}.

Note that the only difference between the configurations $\mathbb{S}_1$ and $\mathbb{S}_4$ is a much longer tail in the second case ($N=8000$ instead of the original $N=4000$). In the figure, we can see that for both the shown cases there is an established quasi steady-state region.
However, for a shorter chain where $N=4000$, the instantaneous crack speed speed experiences a jump at $t_*\approx 3800$. This event indicates the arrival of the reflected wave from the left-hand end of the structure. Despite this phenomenon, the results $v(t_*)$, established before this event, are identical for different $N$, within the accuracy of the evaluation.

\begin{figure}[h!]
\minipage{0.5\textwidth}
\center{\includegraphics[width=\linewidth] {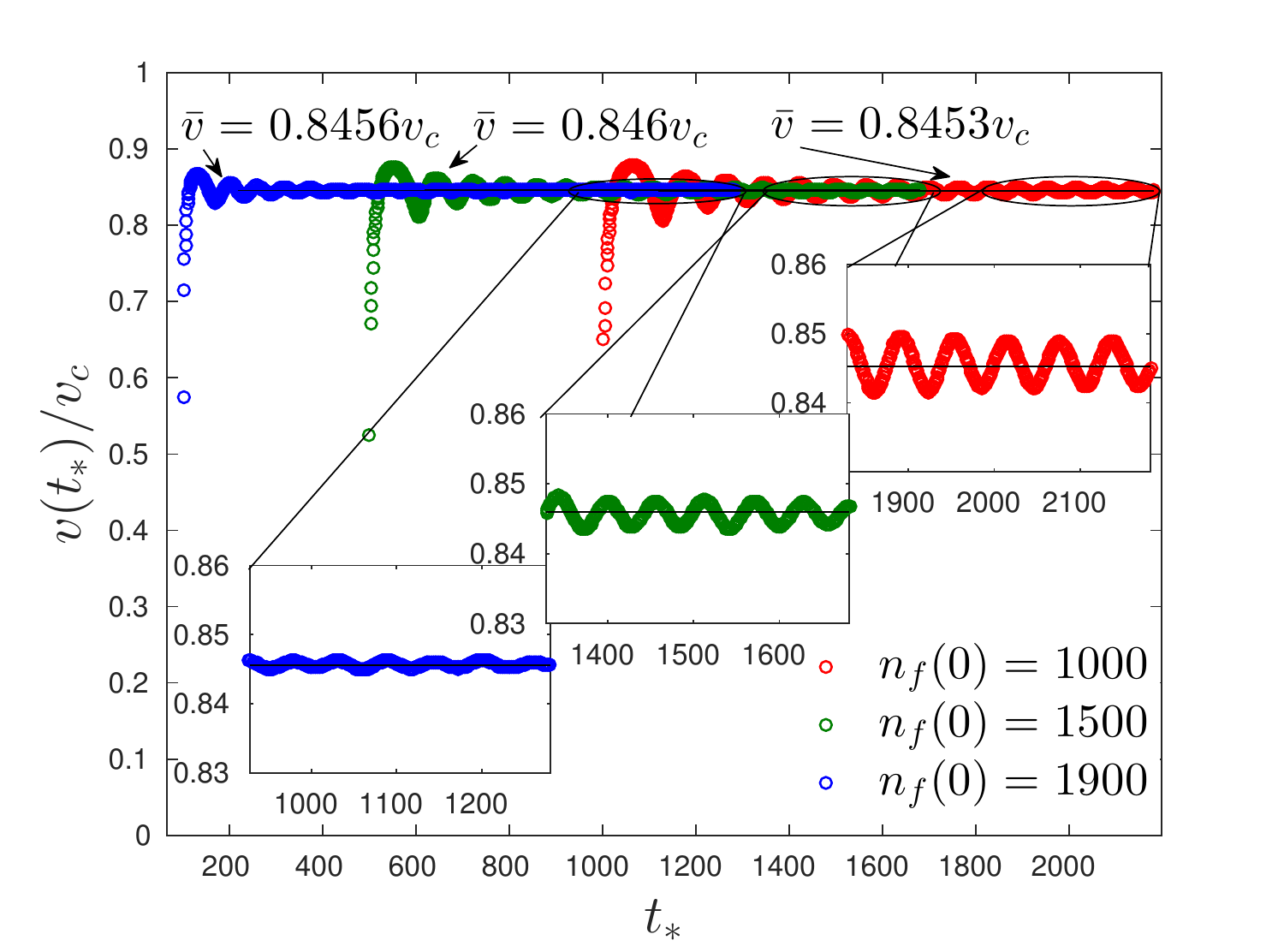} \\ a)}
\endminipage
\hfill
\minipage{0.5\textwidth}
\center{\includegraphics[width=\linewidth] {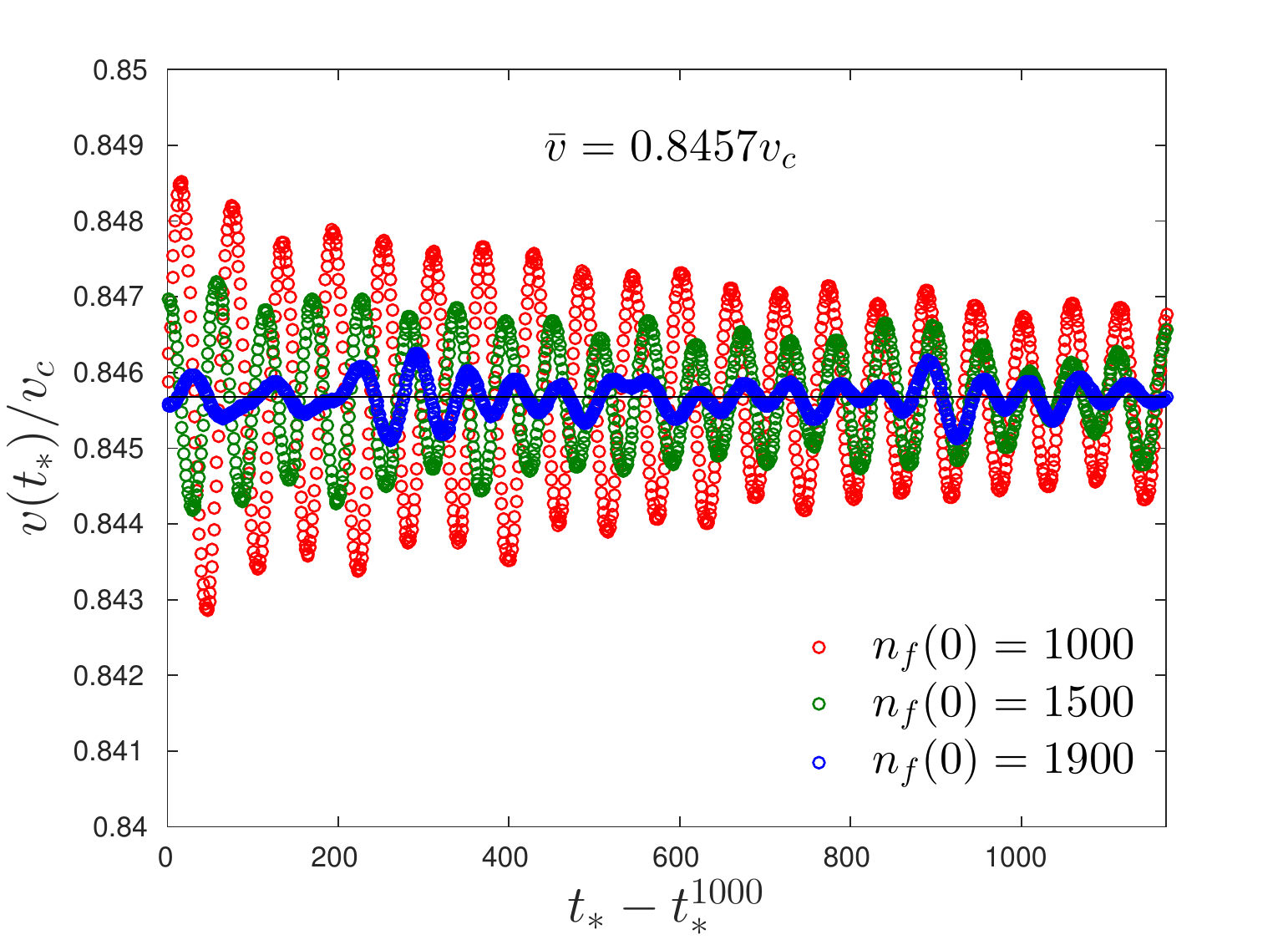} \\ b)}
\endminipage
\captiondelim{. }
\caption[ ]{The Instantaneous crack speed $v(t_*)/v_c$, given by \eqref{eq:CrackSpeed}, for different initial positions of force $n_f(0)$, where $F=5Mu_cv_c^2$, $v_f=0.3v_c$ and $c_1=2c_2$. a) The first $I=1000$ breakages, where the inserts show the final stages of the computations.  The steady-state crack speeds estimated by \eqref{eq:ComputedSpeed} are also shown, and are presented by solid straight lines. b) The continuation of the computations after the 1000th breakage. The estimation of the steady-state crack speed is the same for all the presented cases.}
\label{fig:DifferentNf}
\end{figure}

Finally, we present the effects of different initial distances between the force position $n_f(0)$ and the crack tip $n_*(0)$. We choose the same physical configuration as in the previous subsection, that is $F=5Mu_cv_c^2$, $v_f=0.3v_c$, $c_1=2c_2$ and different geometrical configurations $\mathbb{S}_1$, $\mathbb{S}_2$ and $\mathbb{S}_3$, which give $n_f(0)=1000$, $n_f(0)=1500$ and $n_f(0)=1900$, respectively. The results are shown in Fig. \ref{fig:DifferentNf}a). We can see that the respective steady-state crack speeds calculated from \eqref{eq:ComputedSpeed} are $\bar v=0.8456v_c$, $\bar v=0.846v_c$, $\bar v =0.8453v_c$.
These calculated values of $\bar{v}$ remain within the chosen accuracy up to the third decimal place.
As expected, the fracture process starts earlier for the smaller initial distance between the crack tip and the force position. Moreover, it seems from the computation that the amplitude of the variation of the instantaneous speed, $v_*(t)$, decreases much faster here than in the other two cases, $n_f(0)=1500$ and $n_f(0)=1000$.

This suggests that we can set the initial force location sufficiently close to the crack tip to achieve fast convergence to the desired steady-state regime
and so obtain a more accurate result. However, we avoid this scenario in the computations in our paper for the following simple reason: in the case of a small force moving faster than the crack tip itself, the time interval may become insufficiently long for the cause of making a confident conclusion on the convergence of the process.

As a result of this analysis, and similarly to the case in which we discussed the length of the data sample used in further evaluations, we use configuration $\mathbb{S}_1$ in further computations. This configuration is, in  a sense, worse in comparison to the others when judged on the convergence rate of the fracture process to the steady-state regime. However, since the distance between the crack tip and the load is sufficiently large, it provides more confidence that the analysed phenomena has been properly captured even if we have slightly sacrificed some accuracy and efficiency in the computations.

Finally, in Fig. \ref{fig:DifferentNf}b) we analyse the convergence of the fracture process to a pure steady-state regime, continuing the iterations beyond the chosen limit $I=1000$. Fortunately in this case,  unlike in the example presented in Fig.\ref{fig:DifferentN}, no response from reflected waves deforms the physical picture. We present the corresponding results , starting from the differing moments in time when the fracture processes reach the same link $j=n_*(0)+I$. We may conclude that the processes slowly converge, while the computations for the steady-state crack speed using formula \eqref{eq:ComputedSpeed} using the respective data sample give the consistent value $\bar{v}=0.8457v_c$ which coincides with that predicted analytically.

From the computations performed in this section, we observe that for appropriately chosen geometrical parameters the instantaneous crack speed, $v(t_*)$, stabilises and begins to oscillate about some central value with decreasing amplitude. Even though the full process is not steady-state and, generally speaking, is represented by a transient regime, we can numerically evaluate the average of the crack speed, $\bar v$, and assess the accuracy of the computations. The computations may, however, be time consuming if we want to examine the process for a wide range of structural parameters.

The major question is whether the analytical approach proposed in \cite{slepyan1984}, and developed later for several configurations and load conditions in \cite{ayzenberg2014,slepyan2005,slepyan2012}, may be utilised in the case of the moving load where only a transient regime has been realised, and the steady-state regime is only developed within a region distant from the load and the external boundaries. We are particularly interested in the following questions:
\begin{itemize}
\item Can we predict the behaviour of the local steady-state regime analytically as a function of the prescribed loading (force $F$ and its velocity, $v_f$, if it moves) and the mechanical parameters of the structure?
\item Is it possible to predict both where the crack will move under such conditions and the character of this movement?
\item What is the inclination angle developing behind the crack tip as the result of the fracture process?
\item How does the complete picture of the developed fracture process depend on the problem parameters?
\end{itemize}

In the next sections we first provide the relevant analytical results and then verify them using the numerical computations whose accuracy and robustness we have just verified.

\section{Infinite structure with propagating crack under moving load}
\subsection{Formulation of the problem}
For the theoretical analysis of the problem  described above we consider \emph{an infinite chain} of oscillators, as shown in Fig.\ref{fig:Infinite Chain}. The equations of motion for this system take the same form \eqref{eq:ProblemNumerics},
where the first equation is valid for all $n<n_*(t)$ while the second one corresponds to the intact part of the system and is satisfied on the half axis $n\ge n_*(t)$.
We assume that appropriate radiation conditions at infinity are prescribed. We extensively use the method developed by Slepyan and his co-authors \cite{ayzenberg2014,nieves2016,slepyan2005,slepyan2012,slepyan1984}.

We search for a solution of the problem in a steady-state regime that naturally requires some assumptions for derivation of the final formulae. For the moment, let us assume that at some moment in time the crack speed stabilizes and the crack moves periodically. This means that every breakage occurs within a certain time step and that deformation picture of the entire structure remains (in the moving reference frame coinciding with the crack tip at the moment of breakage) the same at these moments as compared with the equivalent picture at the moment of the previous breakage. In the proceeding analysis we define the time of the beginning of this process as $t=0$.

Following \cite{slepyan2012}, this allows us to introduce a change of variables:
\begin{equation}
\eta=n-n_*(t), \quad n_*(t)=n_f(0)+n_0+vt,
\label{eq:eta}
\end{equation}
where $n_0=n_*(0)-n_f(0)$ is the distance between the crack tip and the force location at the beginning of the steady-state motion,
$v$ is the speed of the moving coordinate system whose origin coincides with the position of the crack tip at moments when breakages occur.
The limitation on the values of crack speed $v$ is given in \eqref{eq:CriticalSpeed}. We assume it is a known parameter whose value remains to be determined by further analysis.

We introduce a new function:
\begin{equation}
u(\eta,t)=u_{n}(t),
\end{equation}
which depends on two continuous independent variables for any fixed value of $n$.
In the moving coordinate system, the equations of motion \eqref{eq:ProblemNumerics} for the new function is written in the form:
\begin{equation}
\begin{gathered}
M\left(\frac{\partial^2}{\partial t^2}-2v\frac{\partial^2}{\partial t\partial \eta} +v^2\frac{\partial^2}{\partial\eta^2}\right)u(\eta,t)=\\
c_2(u(\eta+1,t)+u(\eta-1,t)-2u(\eta,t))
-2c_1u(\eta,t)H(\eta)+F\delta(\eta+n_0+(v-v_f)t),
\label{eq:ProblemInitialEta}
\end{gathered}
\end{equation}
where $H(\eta)$ is the Heaviside step function, and $\delta(\eta)$ is Dirac delta function. As we changed variables in \eqref{eq:eta}, we also modify the derivative with respect to time, which has been incorporated into  \eqref{eq:ProblemInitialEta}.
The initial conditions for this new formulation become:
\begin{equation}
u(\eta,t)=f_0(\eta),\quad \left(\frac{\partial}{\partial t}-v\frac{\partial}{\partial\eta}\right)u(\eta,t)=g_0(\eta),\quad t=0,
\end{equation}
where $f_0$ and $q_0$ are unknown and unimportant functions, since we are here concentrating our efforts on the analysis of a possible steady-state solution.

The subsequent application of Fourier and Laplace transforms to equation \eqref{eq:ProblemInitialEta} reduces it to:
\begin{equation}
\left[(s+ikv)^2+\omega_1^2(k)\right]U^+(k,s)+\left[(s+ikv)^2+\omega_2^2(k)\right]U^-(k,s)=
\frac{Fe^{-ikn_0}}{M}\frac{1}{s+ik(v-v_f)}+H_0(k),
\label{eq:LaplaceFourierTransform}
\end{equation}
where the last term $H_0(k)$ encapsulates the initial conditions. Meanwhile, the  functions $\omega_{1}^2(k)$ and $\omega_{2}^2(k)$
\begin{equation}
\omega_1^2(k)=\omega_2^2(k)+\omega_0^2,\quad \omega_2^2(k)=\frac{4c_2}{M}\sin^2{\left(\frac{k}{2}\right)}, \quad \omega_0^2=\frac{c_1}{M},
\label{eq:DispersionRelations}
\end{equation}
characterise the dispersion relations of the destroyed and intact parts of the structure, respectively, and thus define possible scenarios for wave propagation.

The unknown functions $U^{\pm}(k,s)$ are analytic in the respective half-planes $\pm \Im (k)>0$, and defined as follows:
\begin{equation}
\begin{gathered}
U(k,s)=\int_{0}^{\infty}\left[\int_{-\infty}^{\infty}u(\eta,t)e^{ik\eta}d\eta\right]\,e^{-st}dt=U^+(k,s)+U^-(k,s),\\
U^{\pm}(k,s)=\int_{0}^{\infty}\left[\int_{-\infty}^{\infty}u(\eta,t)H(\pm\eta)e^{ik\eta}d\eta\right]\,e^{-st}dt.
\end{gathered}
\end{equation}
Equation \eqref{eq:LaplaceFourierTransform} can be written in the form of the inhomogeneous Wiener-Hopf equation:
\begin{equation}
L(k,s)U^+(k,s)+U^-(k,s)=\frac{Fe^{-ikn_0}}{M}\frac{1}{s+ik(v-v_f)}\frac{1}{(s+ikv)^2+\omega_2^2(k)}+\frac{H_0(k)}{(s+ikv)^2+\omega_2^2(k)}.
\label{eq:WienerHopf_Initial}
\end{equation}
with the kernel function $L(k,s)$:
\begin{equation}
L(k,s)=\frac{(s+ikv)^2+\omega_1^2(k)}{(s+ikv)^2+\omega_2^2(k)},
\label{eq:FunctionL_with_s}
\end{equation}
One can directly check that for any $s>0$, this function has no zeros along the real axis, $k\in\mathds{R}$, and
possesses the following properties:
\begin{equation}
\begin{gathered}
L(k,s)=\overline{L(-k,s)},\\
|L(k,s)|=|L(-k,s)|,\quad \text{Arg}L(k,s)=-\text{Arg}L(-k,s),\quad \text{for } s>0,\,k\in\mathds{R}.
\end{gathered}
\label{eq:PropertiesL}
\end{equation}
As a result, the kernel has zero index (winding number) \cite{slepyan2012} and is estimated at infinity by the following:
\begin{equation}
L(k,s)=1-\frac{\omega_0^2}{k^2v^2}+O(k^{-4}),\quad  k\to\infty,
\label{eq:Asymptotics_L_infty}
\end{equation}
Utilizing \eqref{eq:PropertiesL} and \eqref{eq:Asymptotics_L_infty}, $L(k,s)$ can be factorised by means of the Cauchy-type integral:
\begin{equation}
L(k,s)=L^+(k,s)L^-(k,s),\quad L^{\pm}(k,s)=\exp{\left(\pm\frac{1}{2\pi i}\int_{-\infty}^{\infty}\frac{\text{Log}L(\xi,s)}{\xi-k}\,d\xi\right)}, \quad \pm \Im k >0.
\label{eq:Factorisation}
\end{equation}
Concerning \eqref{eq:Factorisation}, the Wiener-Hopf equation \eqref{eq:WienerHopf_Initial} reduces to:
\begin{equation}
\begin{gathered}
L^+(k,s)U^+(k,s)+\frac{1}{L^-(k,s)}U^-(k,s)=\\
\frac{Fe^{-ikn_0}}{M}
\frac{1}{s+ik(v-v_f)}\frac{1}{[(s+ikv)^2+\omega_2^2(k)]L^-(k,s)}+
\frac{H_0(k)}{[(s+ikv)^2+\omega_2^2(k)]L^-(k,s)}.
\end{gathered}
\label{eq:WienerHopf_Factorised}
\end{equation}
Taking the right-hand side as the sum of the plus and minus function, we can solve this Wiener-Hopf equation for any fixed value of the variable $s$. Then, inverting both transforms, it is possible to analyse the transient regime of the fracture propagating with a constant speed, $v$. This, however, is rather a computationally challenging task.

Our main interest in the problem considered is to evaluate a possible steady-state solution $u(\eta)$, that is the limit of the function $u(\eta,t)$ as $t\to\infty$:
\begin{equation}
u(\eta)=\lim_{t\to\infty}u(\eta,t)=\lim_{s\to0}s \int_0^{\infty}u(\eta,t)e^{-st}\,dt.
\label{eq:Displacement_limit}
\end{equation}
Here, the second relationship follows from the finite value theorem for Laplace transform, where we assume that the limits in \eqref{eq:Displacement_limit} exist.
It is noted in \cite{slepyan2012} that the existence of the limit is equivalent to the causality principle.
The fracture criterion in \eqref{eq:FractureCondition} for the steady-state regime becomes:
\begin{equation}
\begin{gathered}
u(0)=u_c,\\
u(\eta)<u_c,\quad \eta>0.
\end{gathered}
\label{eq:FractureCondition_eta}
\end{equation}

\subsection{Evaluation of the limiting steady-state regime}

To find the steady-state solution, we multiply the Wiener-Hopf equation \eqref{eq:WienerHopf_Factorised} by $s$ and pass it to the limit $s\to0+$ to obtain
(see Supplementary material for technical details):
\begin{equation}
L^+(k)U^+(k)+\frac{1}{L^-(k)}U^-(k)=\frac{C}{0-ik}+\frac{C}{0+ik},
\label{eq:WienerHopf_Final}
\end{equation}
where
\begin{equation}
U(k)=\lim_{s\to0+}sU(k,s),\quad U^{\pm}(k)=\lim_{s\to0+}sU^{\pm}(k,s),\\
\label{eq:Limit_Fourier_FunctionL}
\end{equation}
\begin{equation}
L(k)=\lim_{s\to0+}L(k,s)=\frac{(0+ikv)^2+\omega_1^2(k)}{(0+ikv)^2+\omega_2^2(k)},
\quad L^{\pm}(k)=\lim_{s\to0+}L^{\pm}(k,s).
\label{eq:FunctionL}
\end{equation}
The expressions $(0\pm ikv)$ should be understood as follows:
\begin{equation}
(0\pm ikv)=\lim_{s\to0+}(s\pm ikv).
\label{eq:Limit}
\end{equation}
The constant $C$ in \eqref{eq:WienerHopf_Final} follows from the analysis of the right-hand side of \eqref{eq:WienerHopf_Factorised} and is summarised in equation (SM18) in the Supplementary material:
\begin{equation}
C=\frac{F}{M}\frac{v_c-v}{v_c-v_f}\frac{R}{\sqrt{\omega_0^2(v_c^2-v^2)}}.
\label{eq:Constant_C}
\end{equation}
The auxiliary parameter $R$ in the last expression is related to the energy balance of the system  and plays a crucial role in the further analysis:
\begin{equation}
R=R(v)=\exp{\left(\frac{1}{\pi}\int_{0}^{\infty}\frac{\text{Arg}L(k)}{k}\,dk\right)}.
\label{eq:function_R}
\end{equation}
The asymptotic behaviours of the factors $L^{\pm}(k)$ give:
\begin{equation}
L^{\pm}(k)=1\pm i\frac{Q}{k}+O\left(\frac{1}{k^2}\right),\quad k\to\infty,\quad
Q=\frac{1}{\pi}\int\limits_{0}^{\infty}\text{log}{|L(\xi)|}\,d\xi,
\label{eq:Asymptotics_L+-_infty}
\end{equation}
\begin{equation}
L^{\pm}(k)= \frac{\omega_0}{\sqrt{v_c^2-v^2}}\frac{R^{\pm1}}{0\mp ik}\left(1+(0\mp ik)S\right)+O(k),\quad k\to0,\quad
\quad S=\frac{1}{\pi}\int\limits_0^\infty\frac{\log{|L(\xi)|}}{\xi^2}\,d\xi.
\label{eq:Asymptototics_L+-_zero}
\end{equation}
Let us observe that the displacement field is expected to be continuous in the vicinity of the crack tip $\eta=0$ and, hence, that the asymptotics must at least satisfy $U^{\pm}=O(k^{-1})$, $k\to\infty$. The last estimate, together with \eqref{eq:Asymptotics_L+-_infty} allows us to solve the Wiener-Hopf equation \eqref{eq:WienerHopf_Final} by utilising Liouville's theorem:
\begin{equation}
U^+(k)=\frac{C}{0-ik}\frac{1}{L^+(k)},\quad U^-(k)=\frac{C}{0+ik}L^-(k).
\end{equation}
In turn, from \eqref{eq:Asymptotics_L+-_infty} and \eqref{eq:Asymptototics_L+-_zero}, it follows that:
\begin{equation}
\begin{gathered}
U^{\pm}(k)= C\left(\pm\frac{i}{k}+\frac{Q}{k^2}\right)+O(k^{-3}),\quad k\to\infty,\\
U^+(k)= \frac{C}{\omega_0R}\sqrt{v_c^2-v^2}+o(1),\quad k\to0,\\
U^-(k)= \frac{\omega_0C}{R}\frac{1}{\sqrt{v_c^2-v^2}}\left(\frac{1}{(0+ik)^2}+\frac{S}{0+ik}\right)+O(1),\quad k\to0.
\end{gathered}
\label{eq:Asymptotics_Fourier}
\end{equation}
The sought for steady-state solution $u(\eta)$, in terms of the inverse Fourier transform, takes the form:
\begin{equation}
u(\eta)=\frac{1}{2\pi}\int_{-\infty}^{\infty}U^{\pm}(k)e^{-ik\eta}\,dk,\quad \pm\eta>0.
\label{eq:SolutionChainInverseFourier}
\end{equation}
Asymptotic estimates \eqref{eq:Asymptotics_Fourier}, the Abel-Tauber type theorem (Theorem A.5 in \cite{piccolroaz2009}) and Cauchy's residue theorem allow us to obtain the asymptotic behaviour of the solution $u(\eta)$:
\begin{equation}
\begin{gathered}
u(\eta)=C(1-Q\eta)+O(\eta^2),\quad \eta\to0,\\
u(\eta)=-\frac{C}{R}\frac{\omega_0}{\sqrt{v_c^2-v^2}}(\eta-S)+O(1),\quad \eta\to-\infty.
\end{gathered}
\label{eq:inclination}
\end{equation}
We note that the value of the constant part of the leading term of \eqref{eq:inclination}$_2$, as $\eta\to-\infty$, defines the inclination angle of the destroyed part of the structure between the crack tip and the position of the force
(see Fig.\ref{fig:example}b), Fig.\ref{fig:DifferentBoundaryConditions})).
Furthermore, estimate \eqref{eq:Asymptotics_Fourier}$_3$ suggests that there might be oscillations (reflected waves) in the limit $\eta\to\infty$ and that they are included in the $O(1)$ term of \eqref{eq:inclination}$_2$.

The application of fracture condition \eqref{eq:FractureCondition_eta} to \eqref{eq:inclination}$_2$ implies that:
\begin{equation}
C=u_c,
\end{equation}
and, in light of \eqref{eq:Constant_C},
this last result gives the relationship between the loading parameters, $F,v_f$, the geometry of the problem, and the steady-state crack speed, $v$:
\begin{equation}
\frac{F}{Mu_cv_c^2}=\frac{v_c-v_f}{v_c-v}\frac{\omega_0}{Rv_c}\sqrt{\frac{v_c^2-v^2}{v_c^2}}.
\label{eq:Dependence_F_vf_v}
\end{equation}

The latter suggests that, for two different pairs of loading parameters $F^{(1)},v_f^{(1)}$ and $F^{(2)},v_f^{(2)}$ leading to the same steady-state speed, the following is valid:
\begin{equation}
\frac{F^{(1)}}{v_c-v_f^{(1)}}=\frac{F^{(2)}}{v_c-v_f^{(2)}}.
\label{eq:DependenceForDifferent_F_vf}
\end{equation}

\subsection{Analysis of the obtained solution.}

\begin{figure}[h!]
\minipage{0.5\textwidth}
\center{\includegraphics[width=\linewidth] {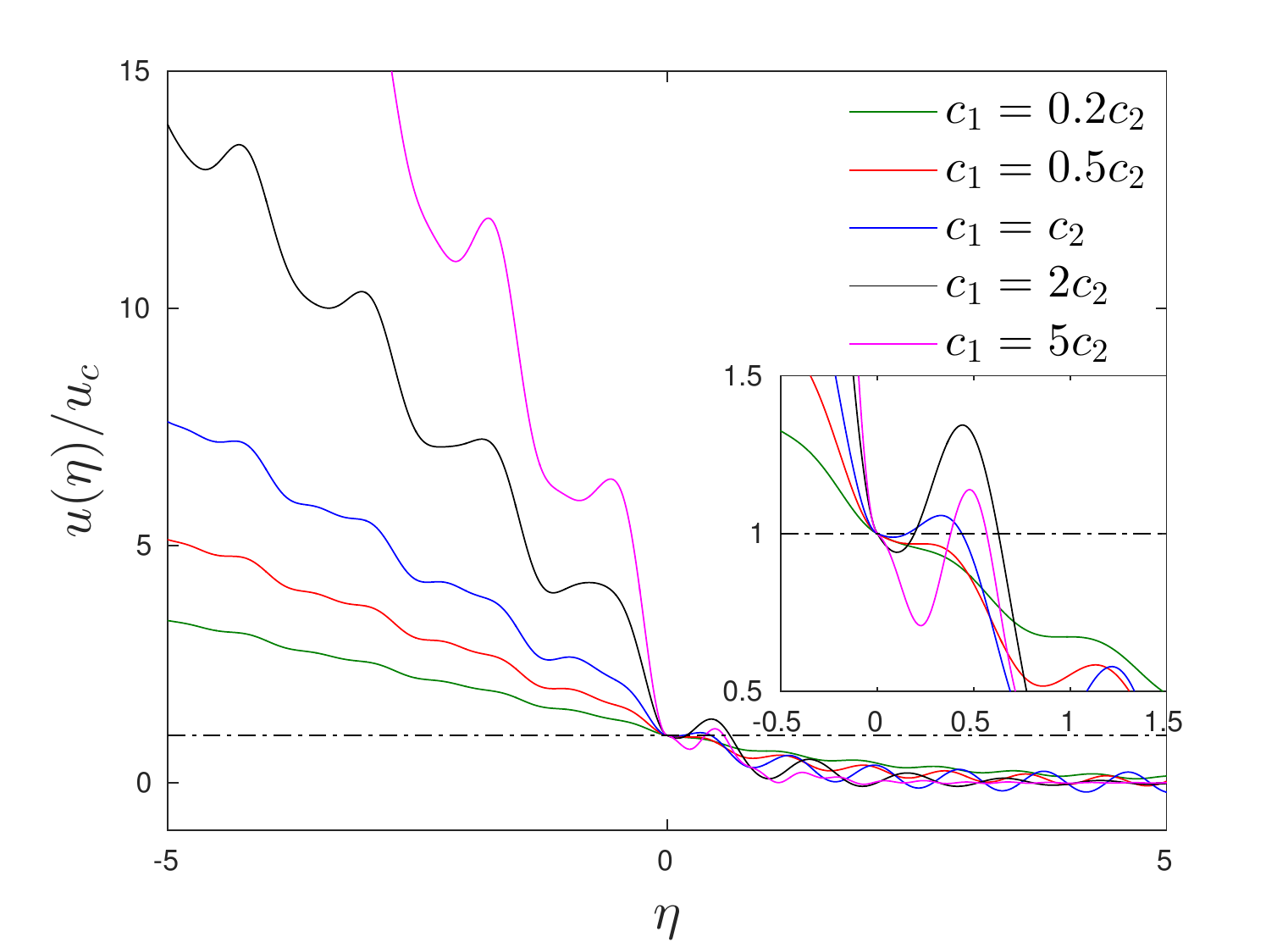} \\ a)}
\endminipage
\hfill
\minipage{0.5\textwidth}
\center{\includegraphics[width=\linewidth] {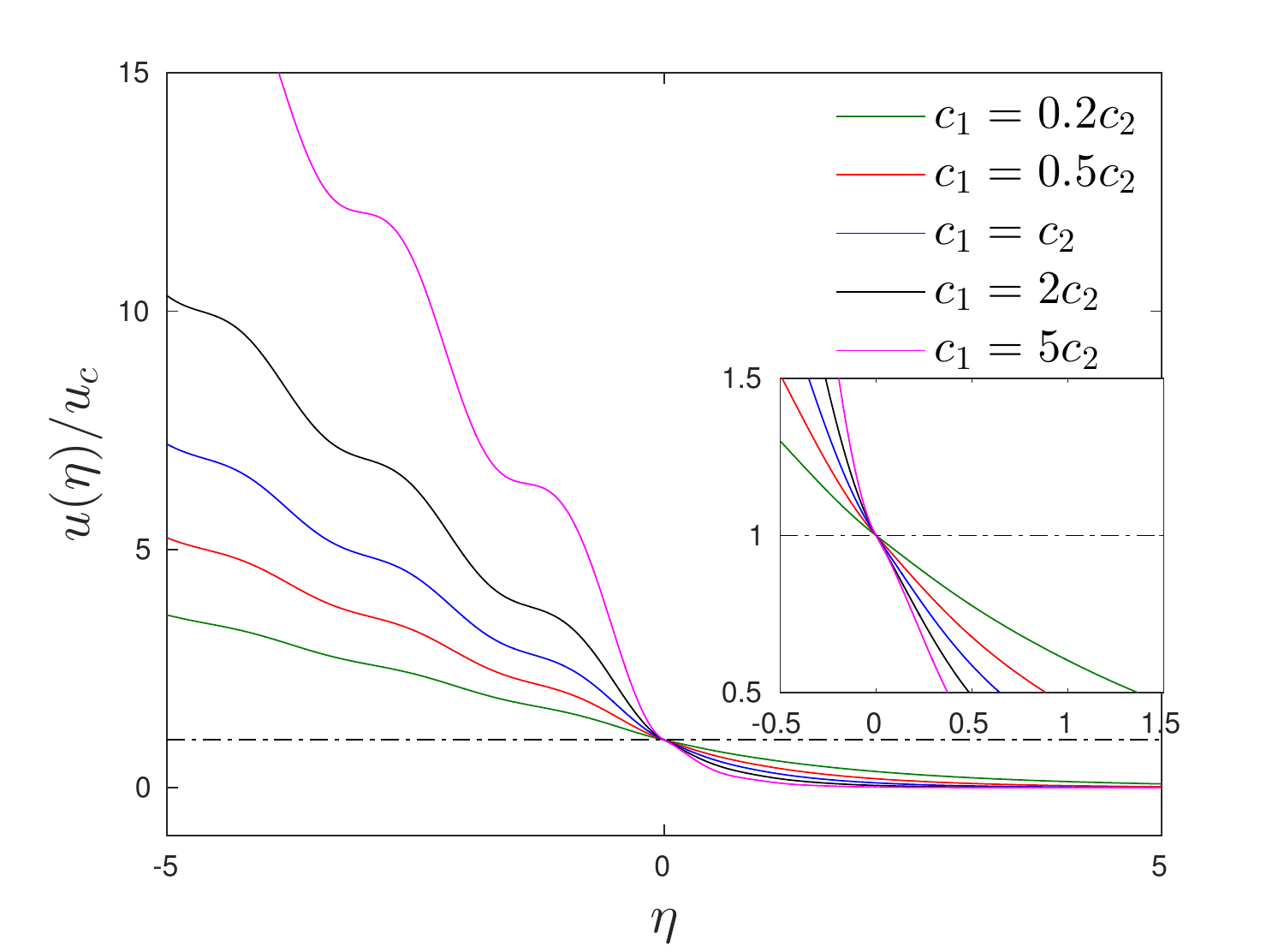} \\ b)}
\endminipage
\captiondelim{. }
\caption[ ]{Displacement field $u(\eta)$ for different values of $c_1$ and different choices of crack speed: a) $v=0.2v_c$, b) $v=0.5v_c$. The inserts show a zoom of the displacement profile in the neighbourhood of the crack tip. The dash-dot line shows the level of displacement $u(\eta)=u_c$. }
\label{fig:Displacements}
\end{figure}

The solution of the problem, $u(\eta)$, is given in terms of the inverse Fourier transform and can be evaluated when a certain crack speed is specified. To illustrate the results, the displacements for the chosen crack speeds are shown in Fig.\ref{fig:Displacements} for different values of $c_1$. In Fig.\ref{fig:Displacements}a), we can see that for $v=0.2v_c$, the second part of fracture condition \eqref{eq:FractureCondition_eta}$_2$ is violated for $c_1/c_2=1,2,5$, whereas for $v=0.5v_c$ in Fig.\ref{fig:Displacements}b) it is fulfilled for every shown case of the stiffness $c_1$.

Following this observation, we can examine the displacement field ahead of the crack tip for every chosen value of $v$ and check the validity of the condition \eqref{eq:FractureCondition_eta}$_2$. A similar analysis was performed for a triangular cell lattice in \cite{kessler1999}. In accordance with fracture condition \eqref{eq:FractureCondition_eta}, the obtained solutions can be divided into two groups:
\begin{itemize}
\item An obtained solution represents an \textbf{admissible regime}
if the fracture condition \eqref{eq:FractureCondition_eta}$_2$ is fulfilled. This regime is fully consistent with the set of assumptions corresponding to the steady-state regime with the given crack speed, $v$.
\item If condition \eqref{eq:FractureCondition_eta}$_2$ is violated, the steady-state propagation regime with speed $v$ is \textbf{forbidden}.
\end{itemize}
Forbidden regimes contain many diverse behaviours, which include clustering \cite{ayzenberg2014} and forerunning \cite{slepyan2015} (also known as a mother-daughter crack mechanism \cite{gao2001}) can be named.

We now analyse the energetic aspect of the  considered problem. The assumptions made on a steady-state regime allow us to introduce local and global energy release rates (ERR) \cite{slepyan2012}. The local ERR, denoted by $G_0$,
corresponds to the potential energy stored in a spring pre-fracture multiplied by the crack speed.
Meanwhile, the global ERR, $G$, characterises the change in energy of the whole structure as the crack moves.
It ensues (see \cite{slepyan1984,slepyan2012}) that the ratio between local ERR $G_0$ and global ERR $G$ is represented by parameter $R$, as defined in \eqref{eq:function_R}:
\begin{equation}
\frac{G_0}{G}=R^2.
\label{eq:ERRRatio}
\end{equation}

We notice that this ratio does not explicitly depend on the loading parameters, and that it was shown in \cite{slepyan2012} that this relation is valid for similar types of loads, such as those constant amplitude, that lead to a steady-state crack propagation.

\begin{figure}[h!]
\center{\includegraphics[scale=0.6]{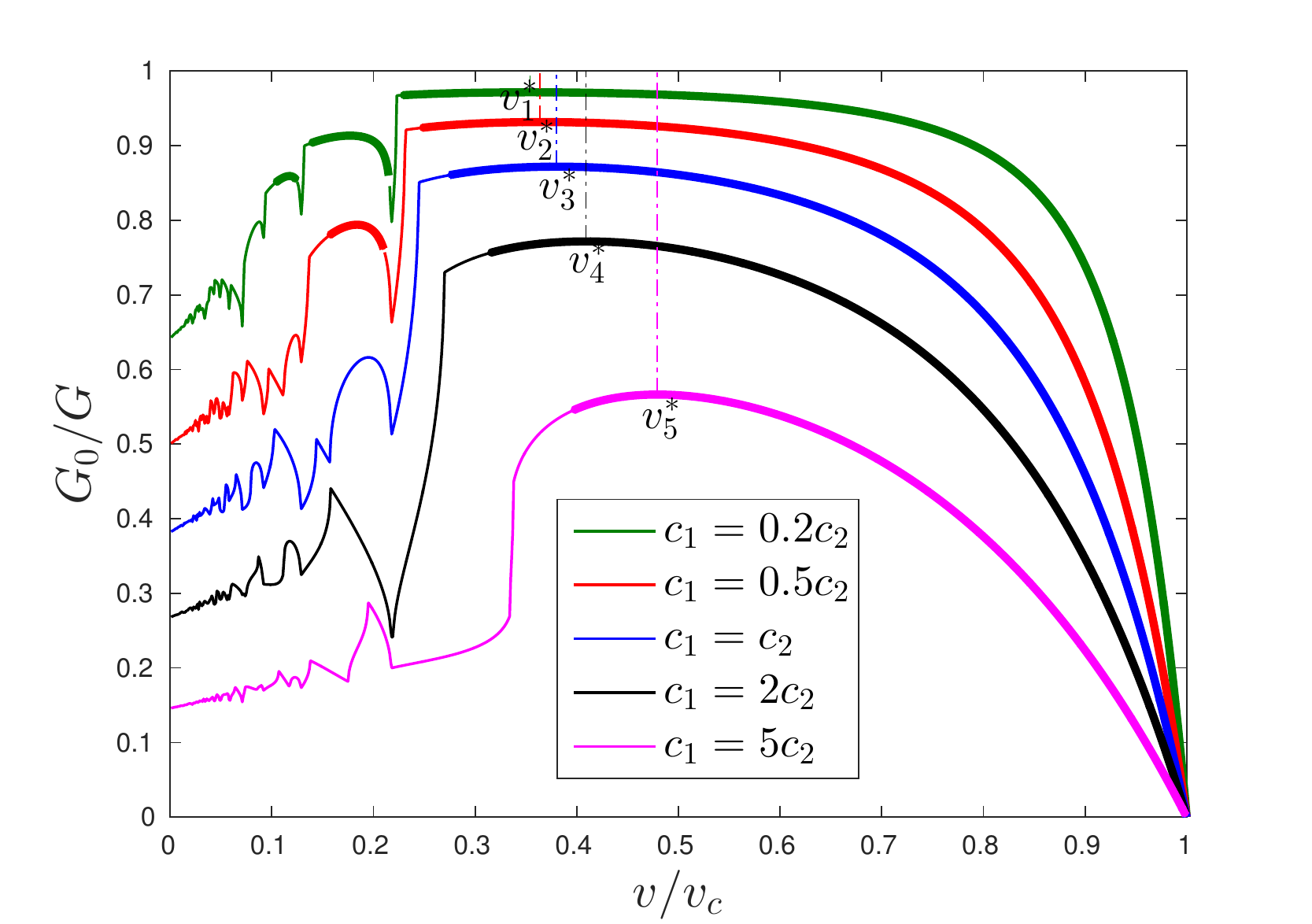}}
\captiondelim{. }
\caption[ ] {Dependence of the ERR ratio $G_0/G$ on the normalised crack speed $v/v_c$ for different values of $c_1/c_2=0.2,0.5,1,2,5$. The specific points $v_j^*$ are the global maximisers of $G_0/G$ and are given in Table \ref{table:TableMaximaERR}. Thick lines correspond to admissible regimes, and thin lines to forbidden ones.}
\label{fig:ERR_Chain}
\end{figure}

The respective energy-speed diagrams are presented in Fig.\ref{fig:ERR_Chain} for five different values of the stiffness $c_1/c_2=0.2,0.5,1,2,5$ of the springs bonding two chains in the intact part of the structure. Since we always assume in this work that $c_2=1$, the varying stiffness $c_1$ defines the anisotropy in the elastic properties of the structure.

In Fig. \ref{fig:ERR_Chain} we also show the crack speeds $v_j^*,j=1,2,...5$, corresponding to the global maxima of $G/G_0$ for various choices of $c_1$. These quantities are also listed in Table \ref{table:TableMaximaERR}.

\begin{table}[!ht]
\begin{center}
\vspace{2mm}
\begin{tabular}{|c|c|c|c|}
\hline
Index $j$ & Value of $c_1/c_2$ & Maximiser $v_j^*/v_c$ & Maximum value of $G_0/G$ \\ \hline
1 & $0.2$ & 0.354 & 0.975 \\ \hline
2 & $0.5$& 0.364 & 0.932 \\ \hline
3 & $1$ & 0.380 & 0.872 \\ \hline
4 & $2$ & 0.409 & 0.772 \\ \hline
5 & $5$ & 0.479 & 0.566 \\
\hline
\end{tabular}
\end{center}
\vspace{-3mm}
\captiondelim{. }
\caption[ ]{The maximisers of $G_0/G$ for different values of $c_1$.}
\label{table:TableMaximaERR}
\end{table}

It should be stressed that similar plots of $G_0/G$ for various structures and loading conditions appear in various papers \cite{ayzenberg2014,marder1995,kessler1999,slepyan1984}. A common feature of these studies is that $G_0/G$ usually possesses a smooth maximum within the intermediate values of $v/v_c$, denoted as $v_j^*,j=1,2,...,5$. It is usually assumed (and commonly agreed) that values $v\geq v_j^*$ are realistic, and that the respective fracture regimes are stable,
while the remaining speeds are assumed to be non-physical, where crack acceleration to the corresponding speed requires as much energy as the estimate  found in the theoretical analysis.
There is no doubt that in such a scenario, the regimes corresponding to speeds in the region $v< v_j^*$ are not admissible. However, a full analysis of the solution reveals that there are physically relevant regimes for $v<v_j^*$. This, consequently, leads to the implication that some values of $G_0/G$ do not correspond to unique and admissible the possible steady state speed, $v$.

This raises a question: in reality, which value for the crack speed is evident on such an occasion?
The theoretical and numerical study of the spontaneous destruction of a discrete structure \cite{ayzenberg2014} has already shown that some regimes of stable crack propagation surely exist, at least for a structure with pronounced anisotropy in its mechanical properties.

From the energy-speed diagram presented in Fig.\ref{fig:ERR_Chain}, we can also conclude that change in the parameter $c_1$ leads to qualitative changes with respect to the number of intervals of admissible regimes. Specifically, there are three distinct intervals for the case $c_1=0.2c_2$, two for the case $c_1=0.5c_2$, and only one for the remaining cases.

\begin{figure}[h!]
\minipage{0.5\textwidth}
\center{\includegraphics[width=\linewidth] {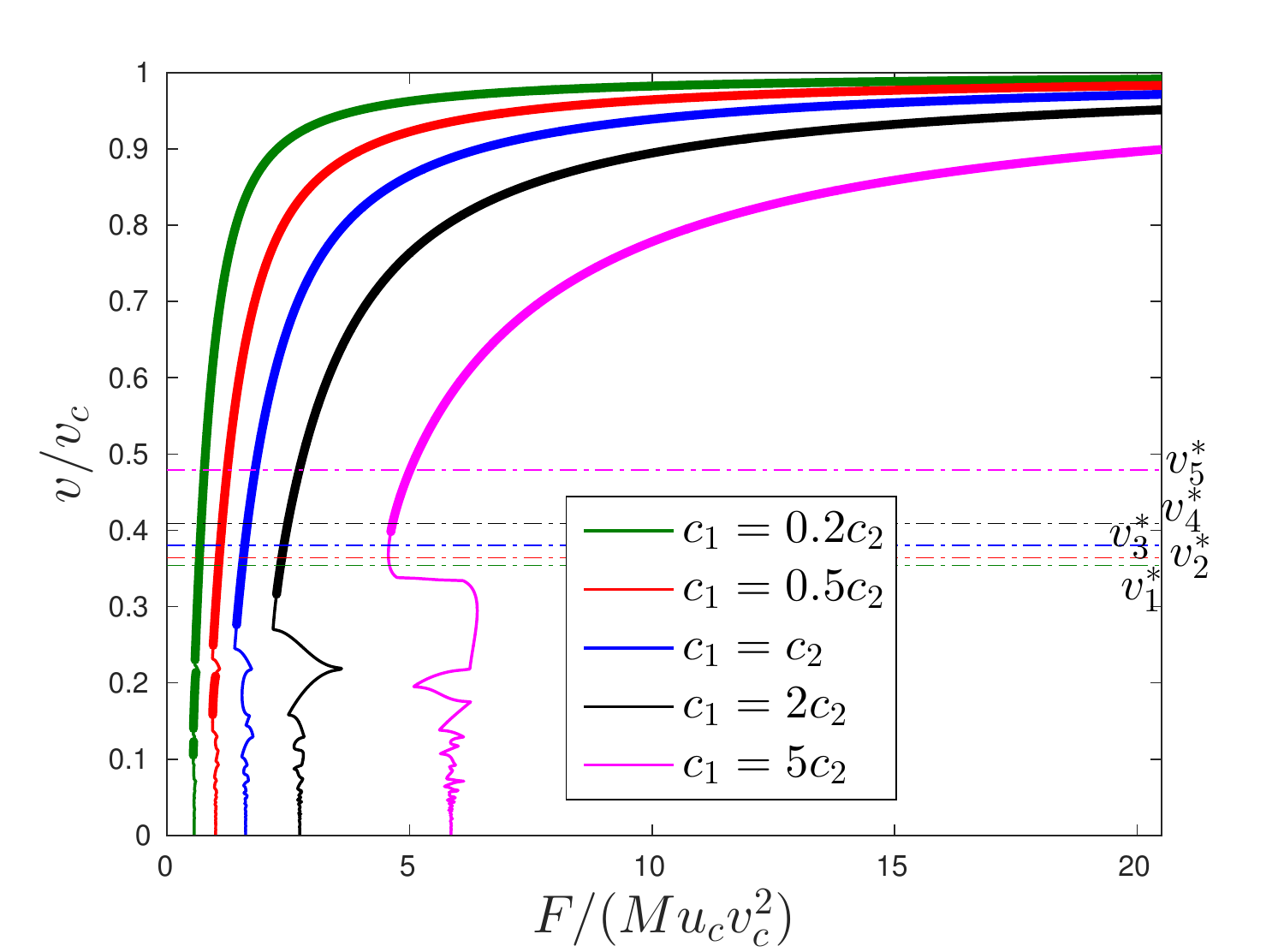} \\ a)}
\endminipage
\hfill
\minipage{0.5\textwidth}
\center{\includegraphics[width=\linewidth] {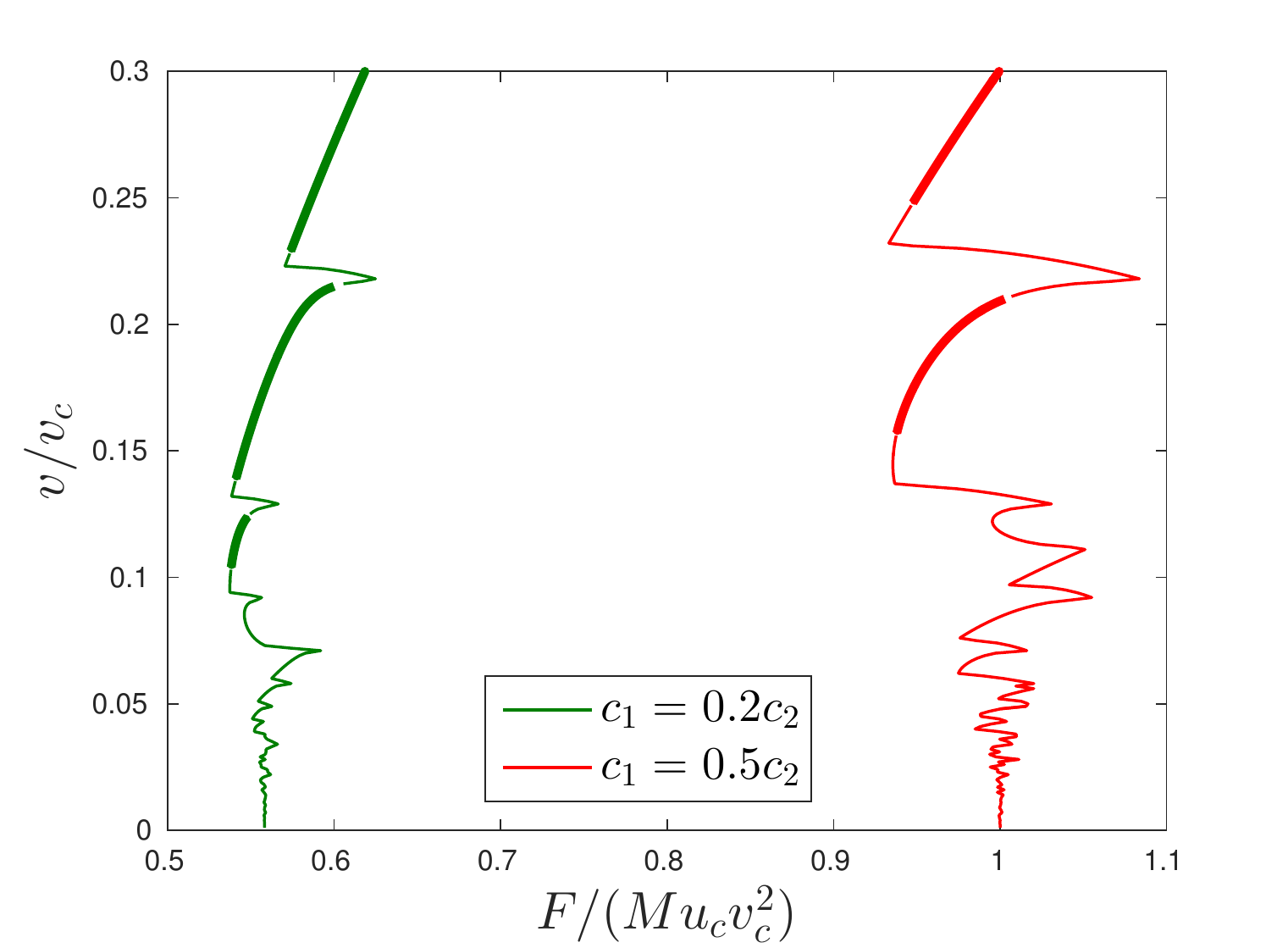} \\ b)}
\endminipage
\captiondelim{. }
\caption[ ]{a) Dependence of crack speed on the normalised force at $v_f=0$ for different values of $c_1$. The dash-dot lines correspond to the maximizers of $G_0/G$ and are given in Table \ref{table:TableData_speed}, b) Zoom of the plot in a) for $c_1/c_2=0.2 c_2, 0.5c_2$ in the region of small values of $v$. Thick lines correspond to the admissible regime, thin lines - forbidden one. }
\label{fig:ERR_Chain_1}
\end{figure}

Along with the energy-speed diagram, we present in Fig.\ref{fig:ERR_Chain_1}a) the dependence of the crack speed, $v$, on the load amplitude, $F$, for fixed load position ($v_f=0$) and several choices of parameter $c_1$, calculated by means of \eqref{eq:Dependence_F_vf_v}. On these plots we again marked the intervals of $v$ corresponding to admissible and forbidden regimes with thick and thin lines, respectively.

The dependence of crack speed on the loading parameter can shed some light on the lack of a unique definition of $v$ for a chosen value of $G_0/G$. Indeed, the plot in
Fig.\ref{fig:ERR_Chain}a) demonstrates the one-to-one correspondence between crack speed and load within the largest regions of admissible regimes. Moreover, the relation \eqref{eq:DependenceForDifferent_F_vf} guarantees that such a correspondence holds true for any choice of $v_f$ and within the period of time that the force location remains behind the crack tip. Simultaneously, this argument shows that a point on the energy-speed diagram may correspond to a set of various combinations of loading parameters $F$ and $v_f$, and thus prompts the question of which point (corresponding steady-state regime) can be accessed at that level of energy, which may be unanswerable without a clear description of the loading history before the steady-state regime is established.

Furthermore, following from Fig.\ref{fig:ERR_Chain}b), a reduction in $c_1$ leads to the appearance of admissible slow steady-state regimes with differing crack speeds for the same value of force, $F$. In other words, even the use of a load-speed diagram does not guarantee the ability to select a unique limiting steady-state crack speed for a pair $F,v_f$. We can, however, see that is impossible for greatly different pairs of $c_1$ and $c_2$.

Another question thus arises: given such non-uniqueness in the diagrams, which choice of crack speed is preferable (if any can be achieved at all) for a given load? Would it be the greatest speed, as commonly accepted?
These questions can not be answered via the discussed analytical model but only by experiments or the direct numerical simulations that we discussed in section \ref{section:NumericalSettings}.

On the other hand, the presented analytical results allow us to compare measurements of crack speed with formulae \eqref{eq:ComputedSpeed} and
\eqref{eq:Dependence_F_vf_v} when a unique solution is present. We can also compare the displacement field $u(\eta)$, as computed from \eqref{eq:SolutionChainInverseFourier} with the one obtained from the numerical simulations.

Notice that the provided loading conditions, i.e. moving force of a fixed magnitude, produce a constant energy flux that causes the crack propagation. In the steady-state crack propagation regime, when $t\to\infty$, the action of the applied force is equivalent to a “remote generalised load”, dependent on the magnitude of original force and its speed. Note also that the speed of the load is lower than the crack speed, thus the steady-state process is observable in a rather large neighbourhood of the crack tip. The work rate, delivered by the applied remote load, balances the dissipation of energy during the fracture (consisting of two parts: the energy necessary for the fracture and the energy of the waves radiated from the crack tip). This fact can be directly shown from the balance of energy fluxes repeating the line of reasoning given in~\cite{slepyan1984}. Finally, this allows us to identify stable admissible regimes when larger values of crack speeds are achieved by the increase of forces. Note however, that in the case of highly anisotropic structure this needs further analysis.

\section{Numerical Simulations and Discussion}

The comparison of the obtained analytical solution with the results of numerical simulations is done in two ways.
We run the numerical simulations for various combinations of loading parameters $F$ and $v_f$ and choices of $c_1$ as described in the previous section. Then, having the respective data for each set of the parameters, we compute the corresponding steady-state crack speed $\bar v$ using formula \eqref{eq:ComputedSpeed}. This allows us to compare the estimated steady-state crack speeds as a function of $F$, $v_f$ with those predicted by the analytical formula \eqref{eq:Dependence_F_vf_v}. As suggested in section \ref{section:NumericalSettings}, most of the computations are performed using the set of geometrical parameters $\mathbb{S}_1$, and cases involving different parameters are explicitly mentioned.

\subsection{Numerical results for isotropic structure}

\begin{figure}[h!]
\minipage{0.5\textwidth}
\center{\includegraphics[width=\linewidth] {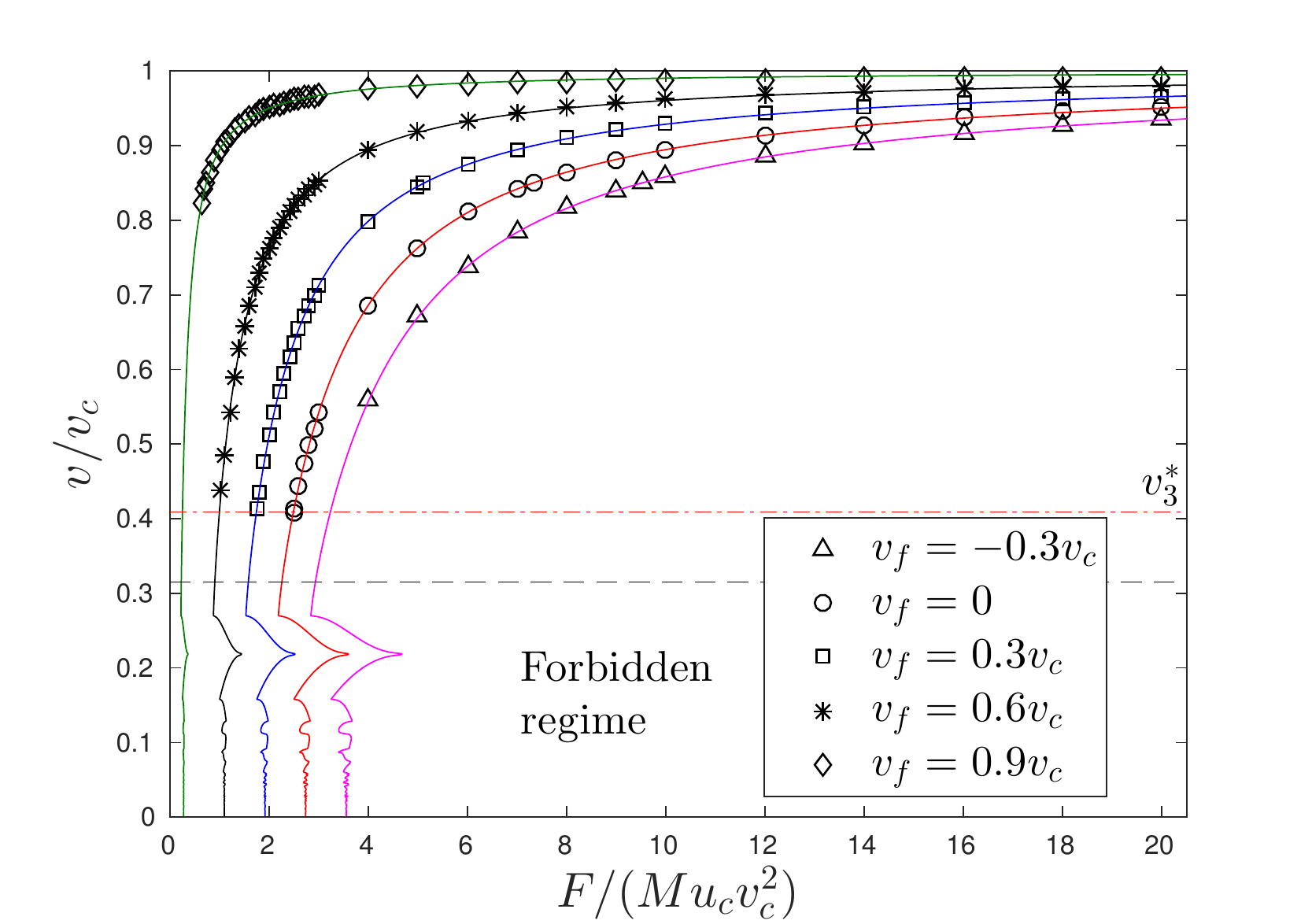} \\ a)}
\endminipage
\hfill
\minipage{0.5\textwidth}
\center{\includegraphics[width=\linewidth] {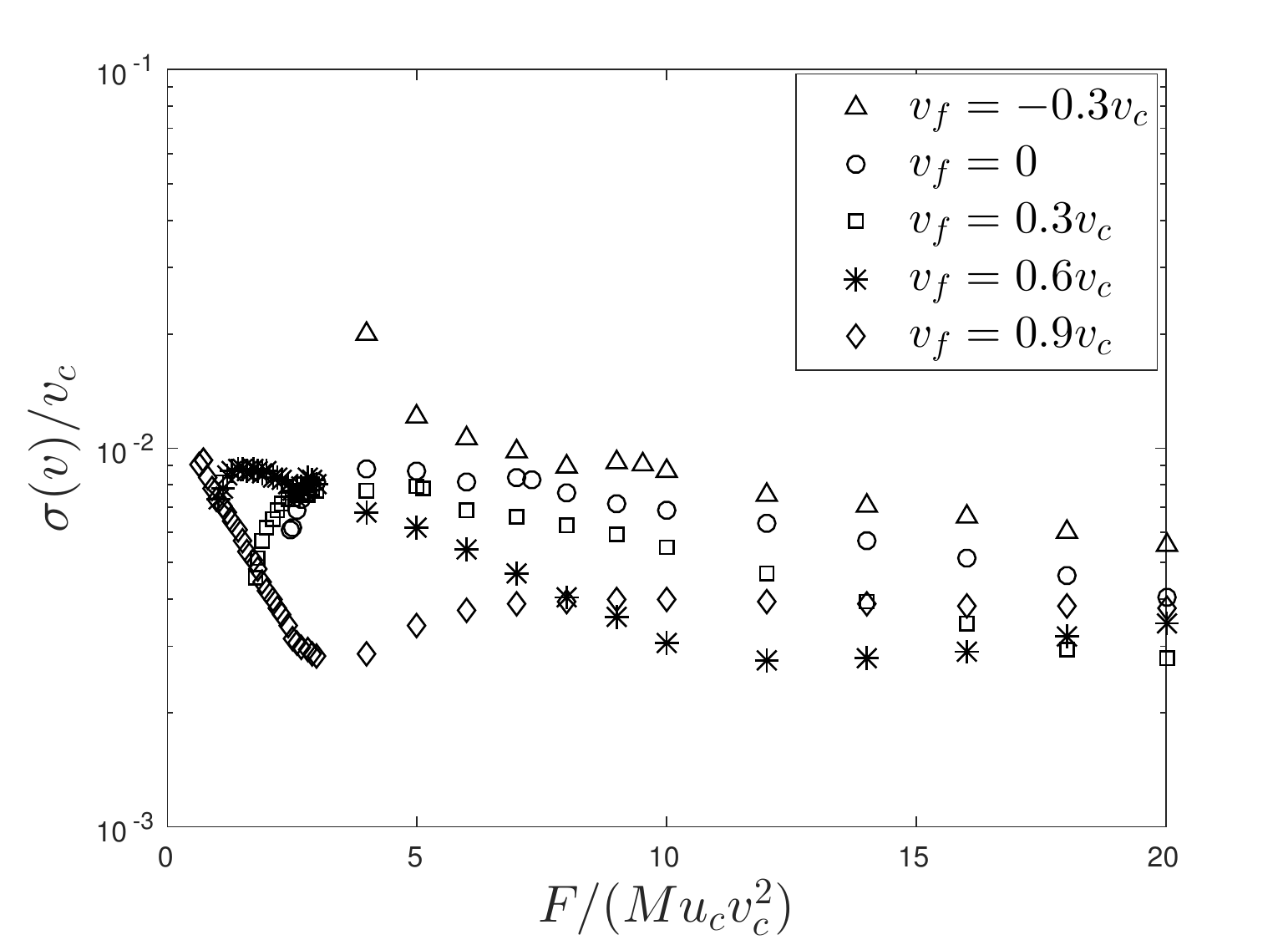} \\ b)}
\endminipage
\captiondelim{. }
\caption[ ]{a) Dependence of the steady-state crack speed for different force magnitude, $F$, and values $v_f/v_c=-0.3,0,0.3,0.6,0.9$ in the isotropic structure case ($c_1=2c_2$). Markers correspond to the results from the numerical simulations, as computed by \eqref{eq:ComputedSpeed}, whereas the solid lines are the calculations made by formula \eqref{eq:Dependence_F_vf_v}. The dash-dotted line shows the value of the maximiser of $G_0/G$ and the dashed line separates the forbidden regime from the admissible one. Fig.\ref{fig:CrackSpeed_Error_vs_Force}b) presents the standard deviation according to \eqref{eq:ComputedSpeed_2}.}
\label{fig:CrackSpeed_Error_vs_Force}
\end{figure}

The results of the numerical evaluation of the steady-state crack speed using equation \eqref{eq:ComputedSpeed} in the case of an isotropic structure, i.e. $c_1=2c_2$, in comparison with the theoretical ones produced from the equation \eqref{eq:Dependence_F_vf_v} are summarised in Fig. \ref{fig:CrackSpeed_Error_vs_Force}a).
Supplementary results showing the standard deviation, calculated using \eqref{eq:ComputedSpeed_2} are given in Fig. \ref{fig:CrackSpeed_Error_vs_Force}b).

Selecting different strengths of the force, $F$, and velocity of its location, $v_f$, we attempt to cover the entire interval of the admissible regime shown in \eqref{eq:ERRRatio} for this case. Numerical results are depicted by markers while their theoretical equivalents are presented by solid lines. Different speeds for the applied force are also considered ($v_f/v_c=-0.3,0,0.3,0.6,0.9$).

The results presented in Fig. \ref{fig:CrackSpeed_Error_vs_Force}a) clearly illustrate that the steady-state regimes predicted by the analytical formulae \eqref{eq:Dependence_F_vf_v} have been attained by the proposed numerical simulations. The results both qualitatively and quantitatively agree within the accuracy estimated by the analysis. Interestingly, by reducing the speed of the force location, $v_f$, we were able to cover a wider region on the energy-speed diagram. However, this strategy has a clear limit as the standard deviation, $\sigma(v)$, demonstrates the opposite behaviour, as is clearly seen from Table.\ref{table:TableSettings}, where in the case of $v_f=-0.3v_c$ we observe a dramatic increase in the standard deviation with reduction in $F$ when using the standard geometrical configuration $\mathbb{S}_1$. For lesser values of the force, we could not identify a clear tendency in the crack propagation regime. This, in turn, makes it impossible to provide a justified comparison between the numerical simulations, indicating that the theory requires further analysis.
Selecting different parameters from Table \ref{table:TableSettings} was unhelpful in the identification of a limiting steady-state regime. An optimal choice which might include the benefits of the aforementioned feature of the numerical process is to consider a fixed force position. In addition to the previous arguments, this would possible small perturbations related to the movement of the force in the numerical simulation.

The minimum achieved steady-state crack speed that we can prove without any doubt did not not become significantly smaller than $v_3^*$ (in Table \ref{table:TableMaximaERR}) for the fixed force ($v_f=0$). On the other hand, the performed theoretical analysis showed a wider range of crack speeds in the admissible region. This observation implies that it may be necessary to take the loading history prior to the steady-state regime into account. More precisely, we should probably take into account how the system reached the value of $G_0/G$ in Fig.\ref{fig:ERR_Chain}.
Moreover, in order to check whether there is anything particular about the points $v_j^*$ we performed simulations for several choices of $c_1$ in the next subsection.

\begin{figure}[ht]
\minipage{0.5\textwidth}
\center{\includegraphics[width=\linewidth] {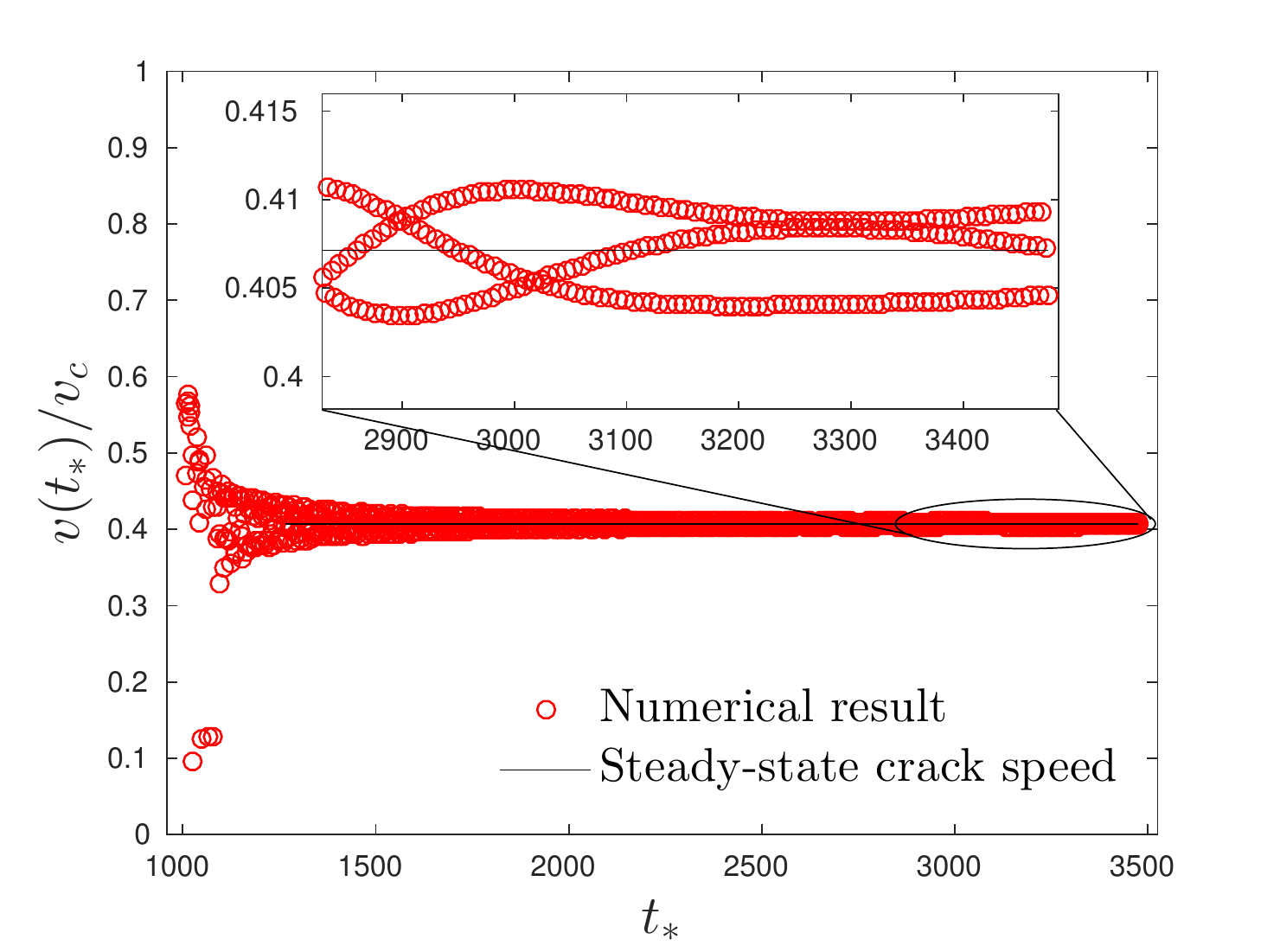} \\ a)}
\endminipage
\hfill
\minipage{0.5\textwidth}
\center{\includegraphics[width=\linewidth]{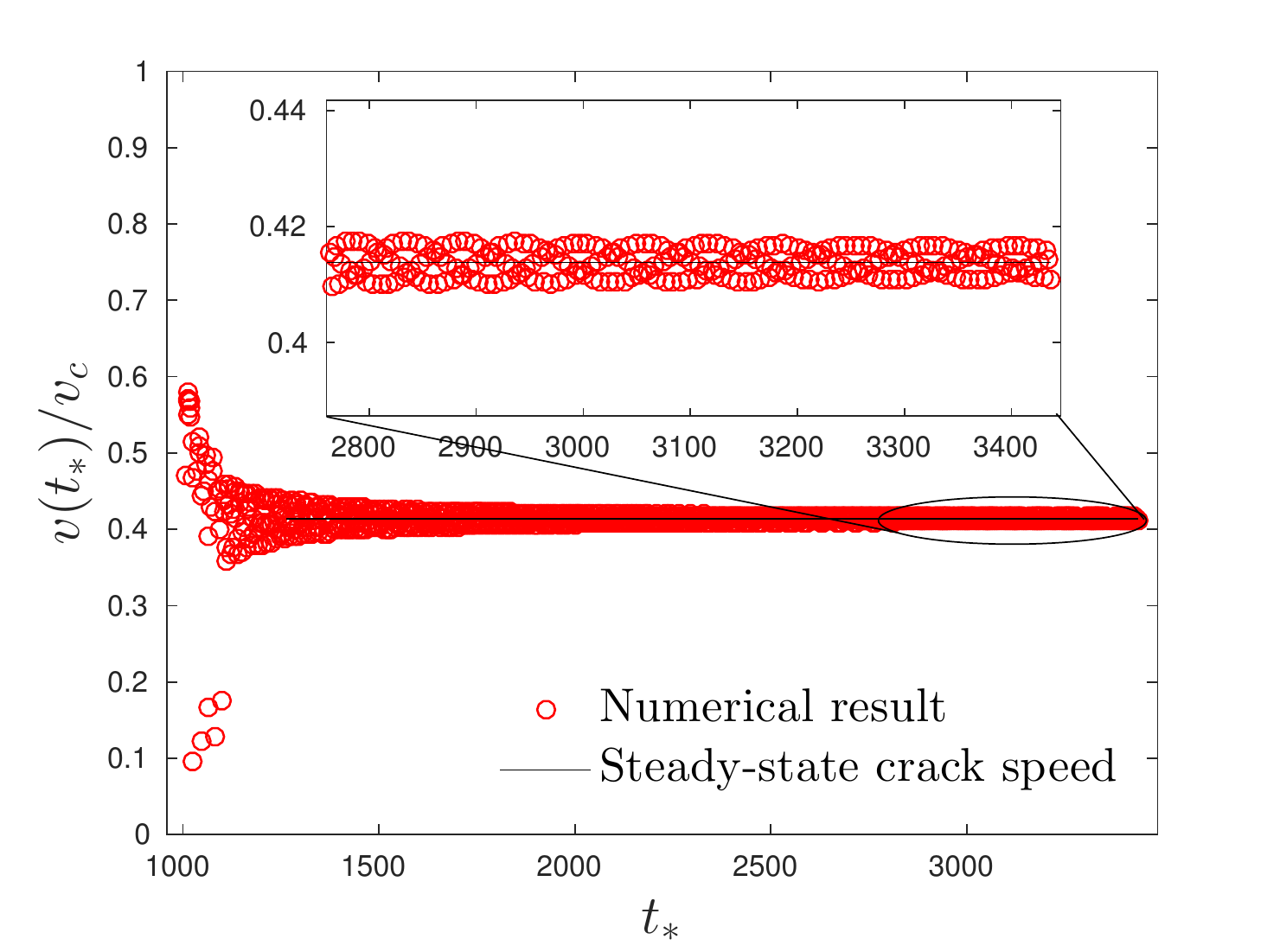} \\ b)}
\endminipage\\
\hfill
\minipage{0.5\textwidth}
\center{\includegraphics[width=\linewidth]{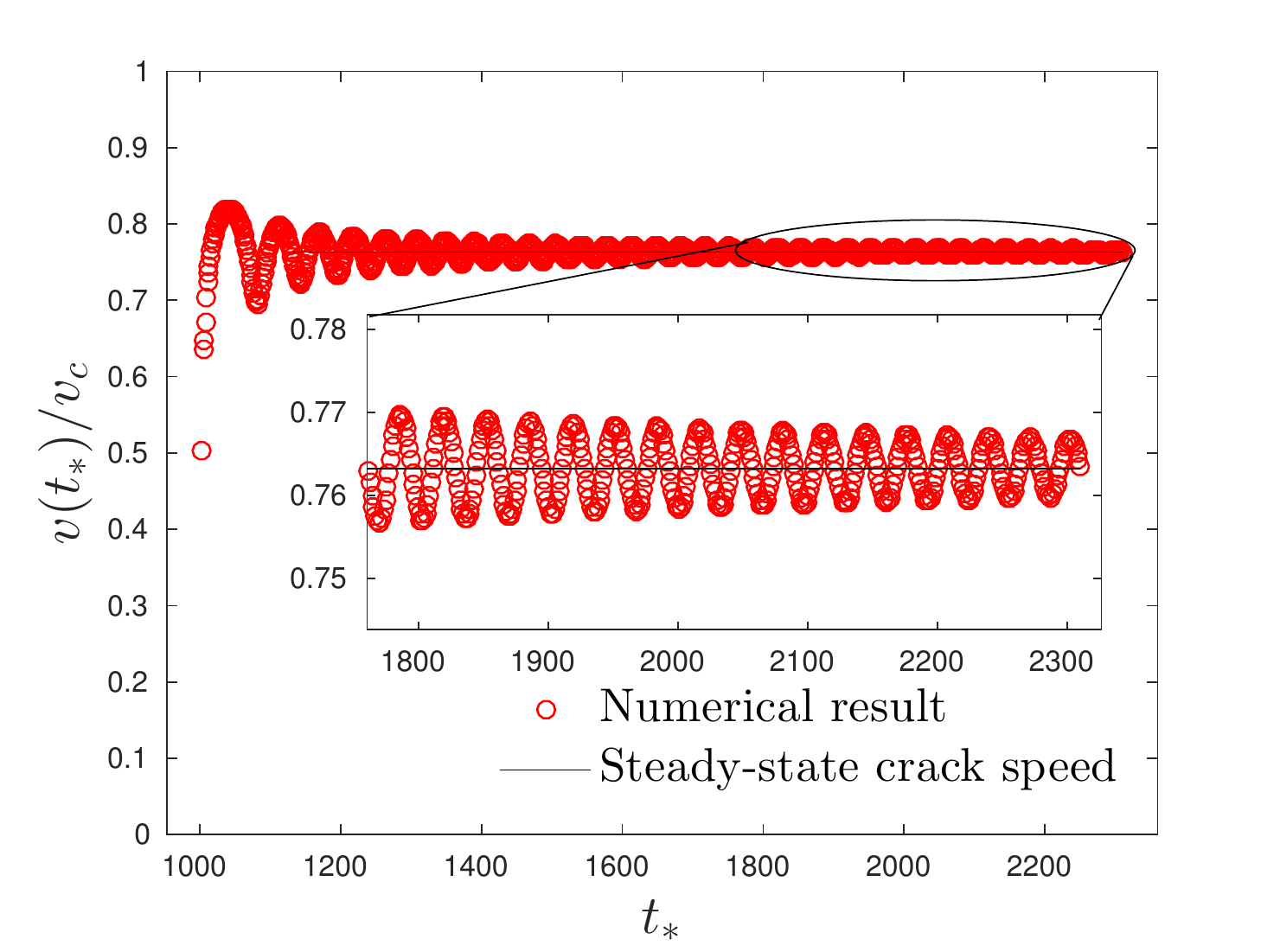} \\ c)}
\endminipage
\hfill
\minipage{0.5\textwidth}
\center{\includegraphics[width=\linewidth]{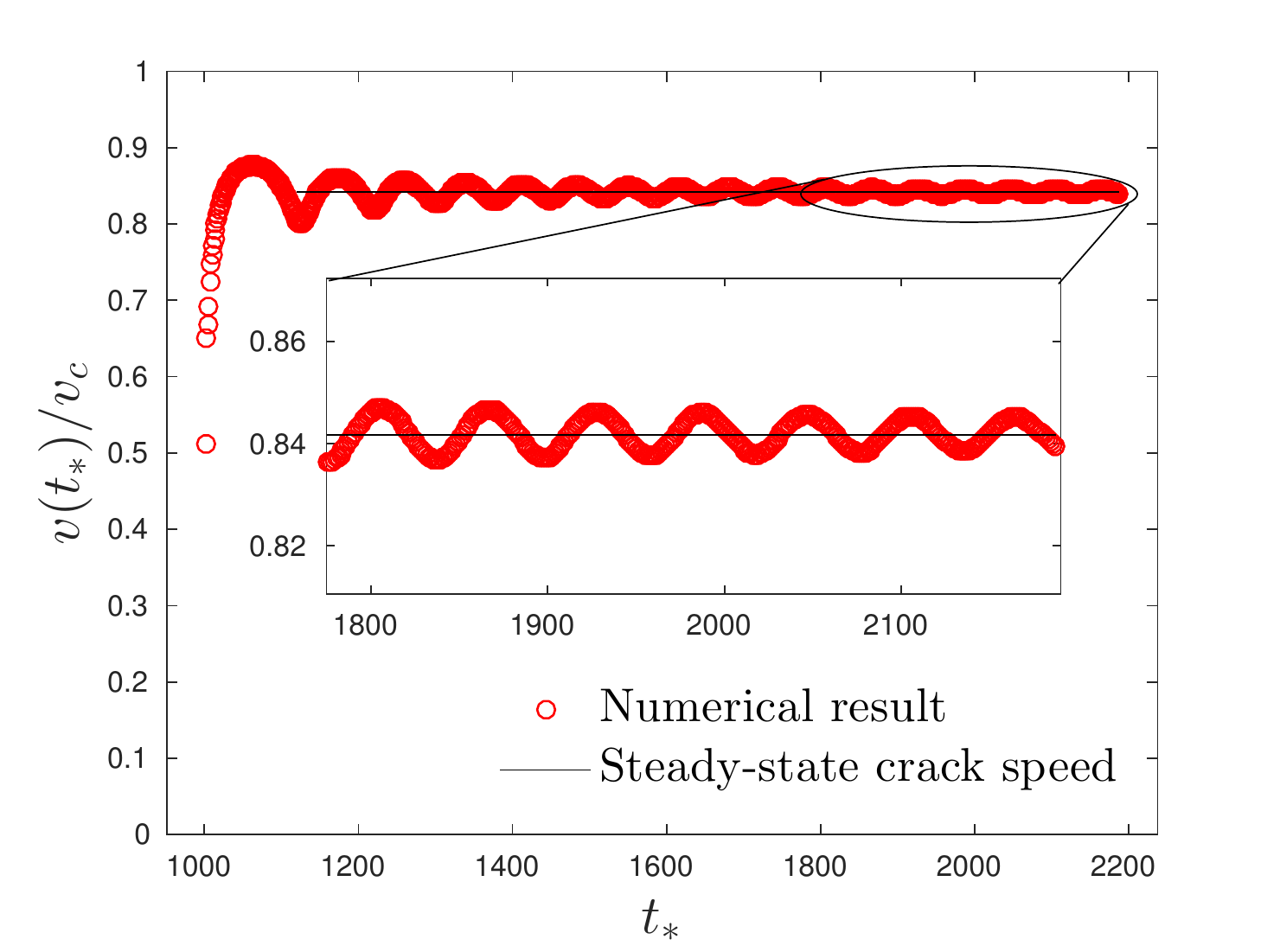} \\ d)}
\endminipage\\
\hfill
\minipage{0.5\textwidth}
\center{\includegraphics[width=\linewidth]{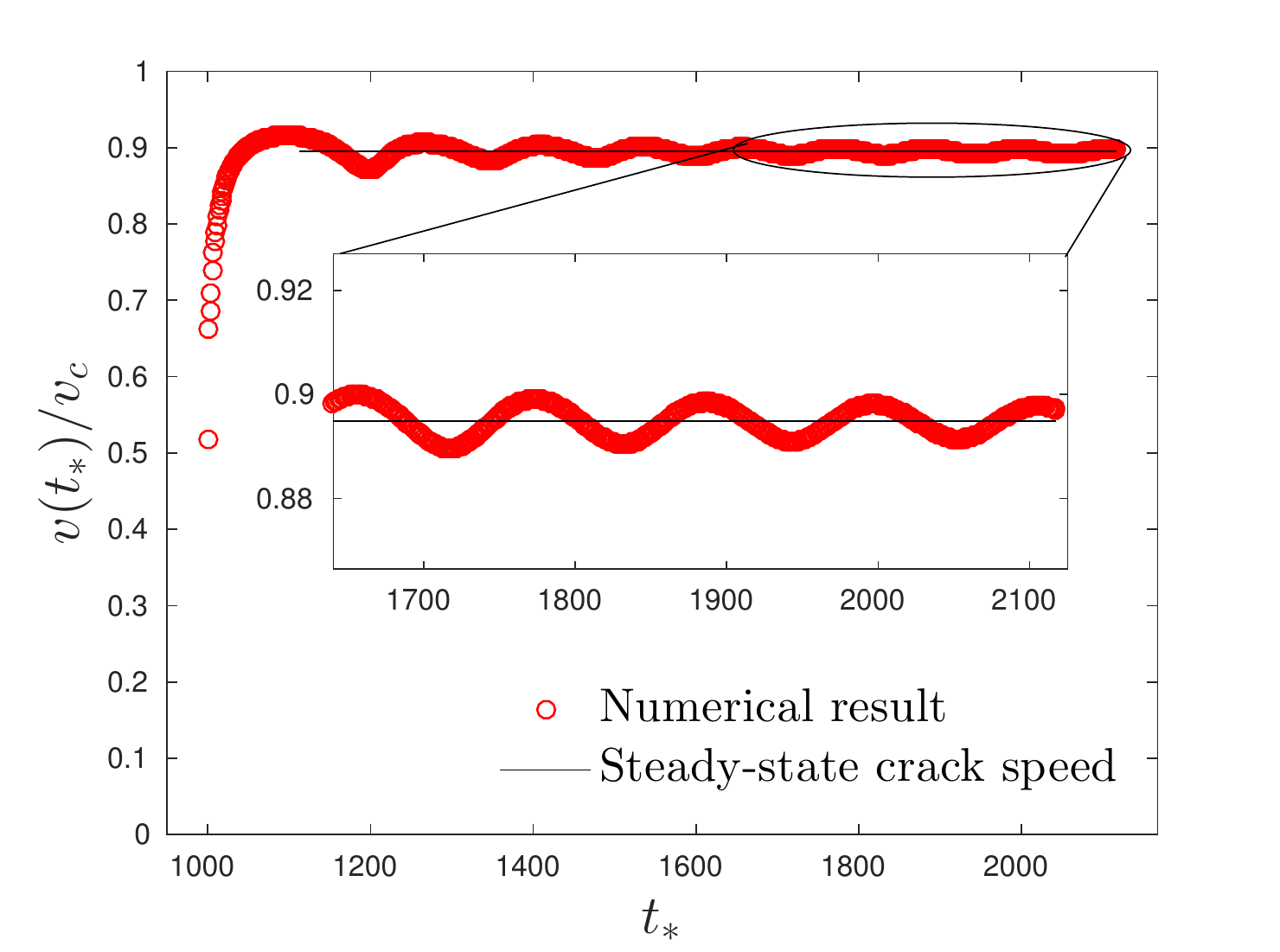} \\ e) }
\endminipage
\hfill
\minipage{0.5\textwidth}
\center{\includegraphics[width=\linewidth]{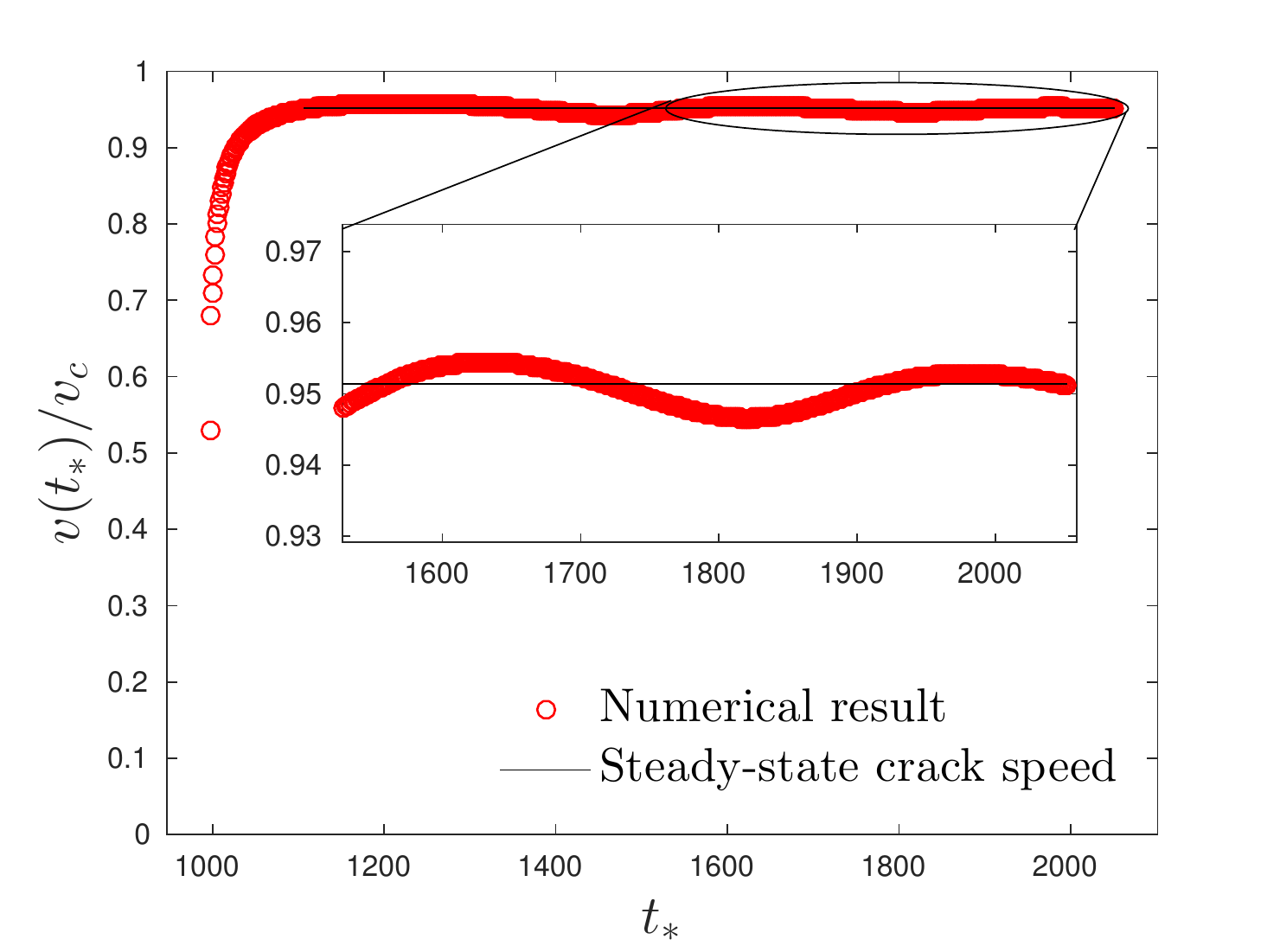} \\ f) }
\endminipage
\captiondelim{. }
\caption[ ]{Instantaneous crack speed $v(t_*)/v_c$ given by \eqref{eq:CrackSpeed} for various strengths of applied force and estimates of steady-state crack speed for the fixed force $v_f=0$: a) $F=2.48Mu_cv_c^2$, $\bar{v}=0.407v_c$, b) $F=2.5Mu_cv_c^2$, $\bar{v}=0.414v_c$, c) $F=5Mu_cv_c^2$, $\bar{v}=0.763v_c$, d) $F=7Mu_cv_c^2$, $\bar{v}=0.842v_c$, e) $F=10Mu_cv_c^2$, $\bar{v}=0.895v_c$, f) $F=20Mu_cv_c^2$, $\bar{v}=0.951v_c$. The inserts demonstrate the dependence of the crack speed during the final stages of the numerical simulations.}
\label{fig:CrackSpeed_vs_Time}
\end{figure}

In Fig.\ref{fig:CrackSpeed_vs_Time} we present distributions of the instantaneous crack speed $v(t_*)$ for various strengths of the force, $F$, when $v_f=0$.
In this set of figures we can observe the behaviour of the instantaneous crack as the amplitude of the force changes. It can be seen that the oscillations of $v(t_*)$ around the corresponding steady-state crack speed are more rapid than for lesser values of $F$. This naturally has a clear effect on the computation of the mean value, $\bar{v}$, from \eqref{eq:ComputedSpeed} and, thus, on the accuracy of the prediction of the steady-state crack speed, $v$.

The observed behaviour also influences the convergence of the transient regime to its steady state. Analysing the behaviour of the instantaneous crack speed for each strength of the force, $F$, allows us to choose the best sample set for use in formulae \eqref{eq:Dependence_F_vf_v}
or may even dictate a need to change the computational configurations (for example, moving the position of the force closer to the crack tip). For the greatest force ($v>0.97v_c$) the character of the force becomes monotonic, while for the weakest force ($v<0.42v_c$) a type of cluster propagation develops.

\begin{figure}[h!]
\minipage{0.5\textwidth}
\center{\includegraphics[width=\linewidth] {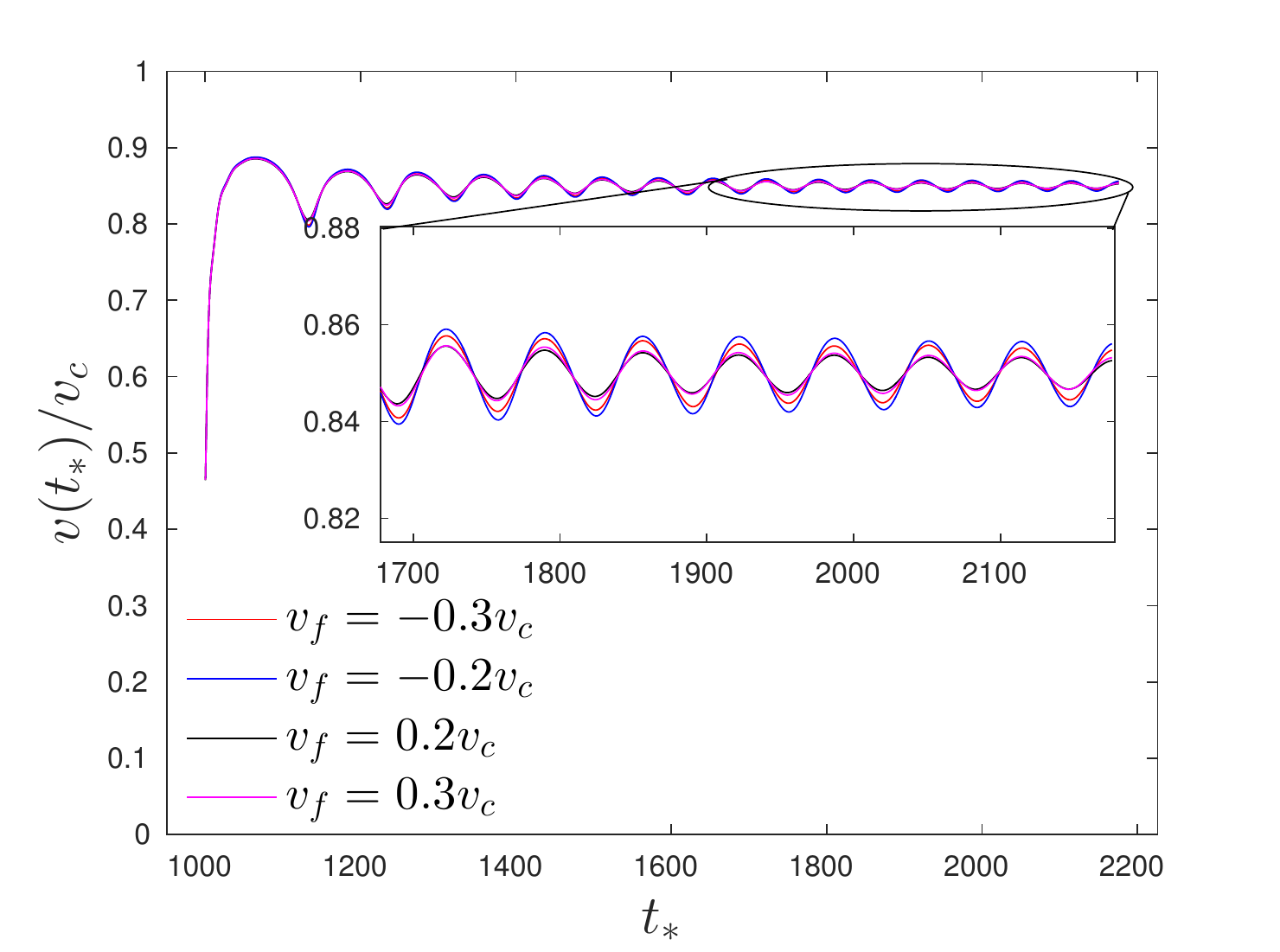} \\ a)}
\endminipage
\hfill
\minipage{0.5\textwidth}
\center{\includegraphics[width=\linewidth]{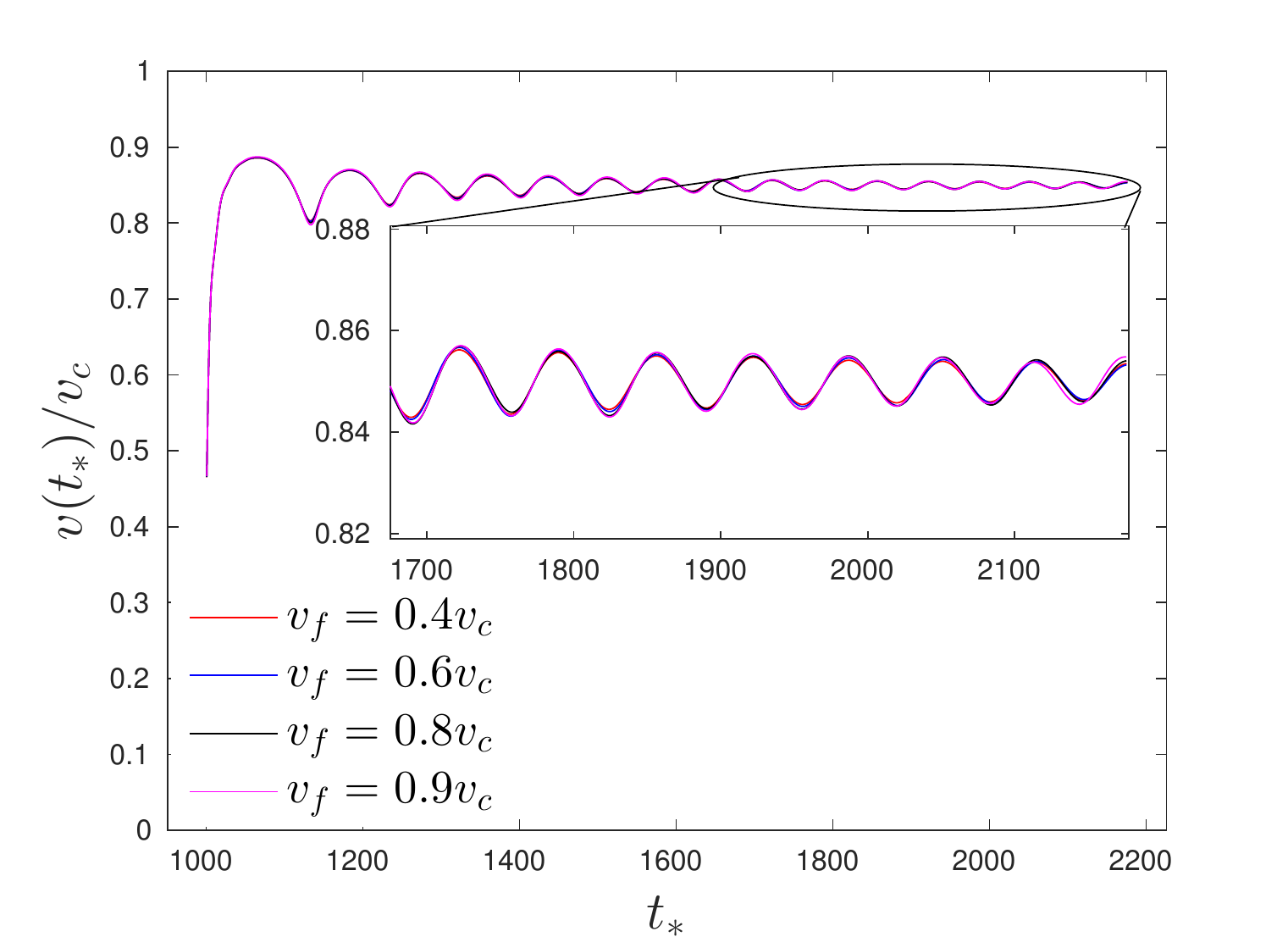} \\ b)}
\endminipage
\captiondelim{. }
\caption[ ]{Time dependence of the normalised crack speed $v(t_*)/v_c$ (smoothed data points) for different choices of $v_f$ and the same estimated $v/v_c=0.85$ and $c_1=2c_2$: a) the force $F/Mu_cv_c^2=9.516,8.784,5.856,5.124$ for $v_f/v_c=-0.3,-0.2,0.2,0.3$, respectively, b) the force $F/Mu_cv_c^2=4.392,2.928,1.464,0.732$ for $v_f/v_c=0.4,0.6,0.8,0.9$, respectively. The inserts demonstrate the dependence of crack speed during the final stages of the numerical simulations. }
\label{fig:SameCrackSpeed}
\end{figure}

Finally, in order to demonstrate the effect of different loading parameters, we consider the behaviour of the instantaneous speed to analyse how (and whether) the process converges to a steady-state regime corresponding to a particular point on the energy-speed diagram. We take a point lying on the stable branch of the energy-speed diagram in Fig.\ref{fig:ERR_Chain} ($v=0.85v_c$).
The plots of the instantaneous crack speed are shown in  Fig.\ref{fig:SameCrackSpeed}a),b)
for various pairs of the loading parameters $F$, $v_f$, leading to the chosen steady-state speed.

\clearpage

Here, we can explicitly observe the accuracy of our numerical simulations as compared to relation \eqref{eq:DependenceForDifferent_F_vf}. The estimate of $\bar{v}$ from \eqref{eq:ComputedSpeed} is the same as the mean value $\bar v$ for all the considered cases of the force speed $v_f$. We separated cases for large (Fig.\ref{fig:SameCrackSpeed}b) and small (Fig.\ref{fig:SameCrackSpeed}a) velocities
of the applied force. For the former, the data correlate perfectly for various speeds, while in the latter, the amplitude becomes sensitive to the
value of the speed, $v_f$. Moreover, the instantaneous crack speed behaves differently in the cases where $|v_f|\leq 0.1v_c$, where no decrease in amplitude of the oscillations in instantaneous crack speed has been observed and the quasi periodicity visible on Fig.\ref{fig:SameCrackSpeed} becomes less pronounced.
However, in spite of this, the computed mean value $\bar{v}$ still matches the predicted value of crack speed from \eqref{eq:DependenceForDifferent_F_vf}.
Thus, a more careful procedure is required to evaluate the limiting steady-state crack speed in the case of a very slow force velocity, $|v_f|<\!<1$.

\subsection{Numerical results for different contrast in elastic properties. Displacement fields.}

\begin{figure}[h!]
\minipage{0.5\textwidth}
\center{\includegraphics[width=\linewidth] {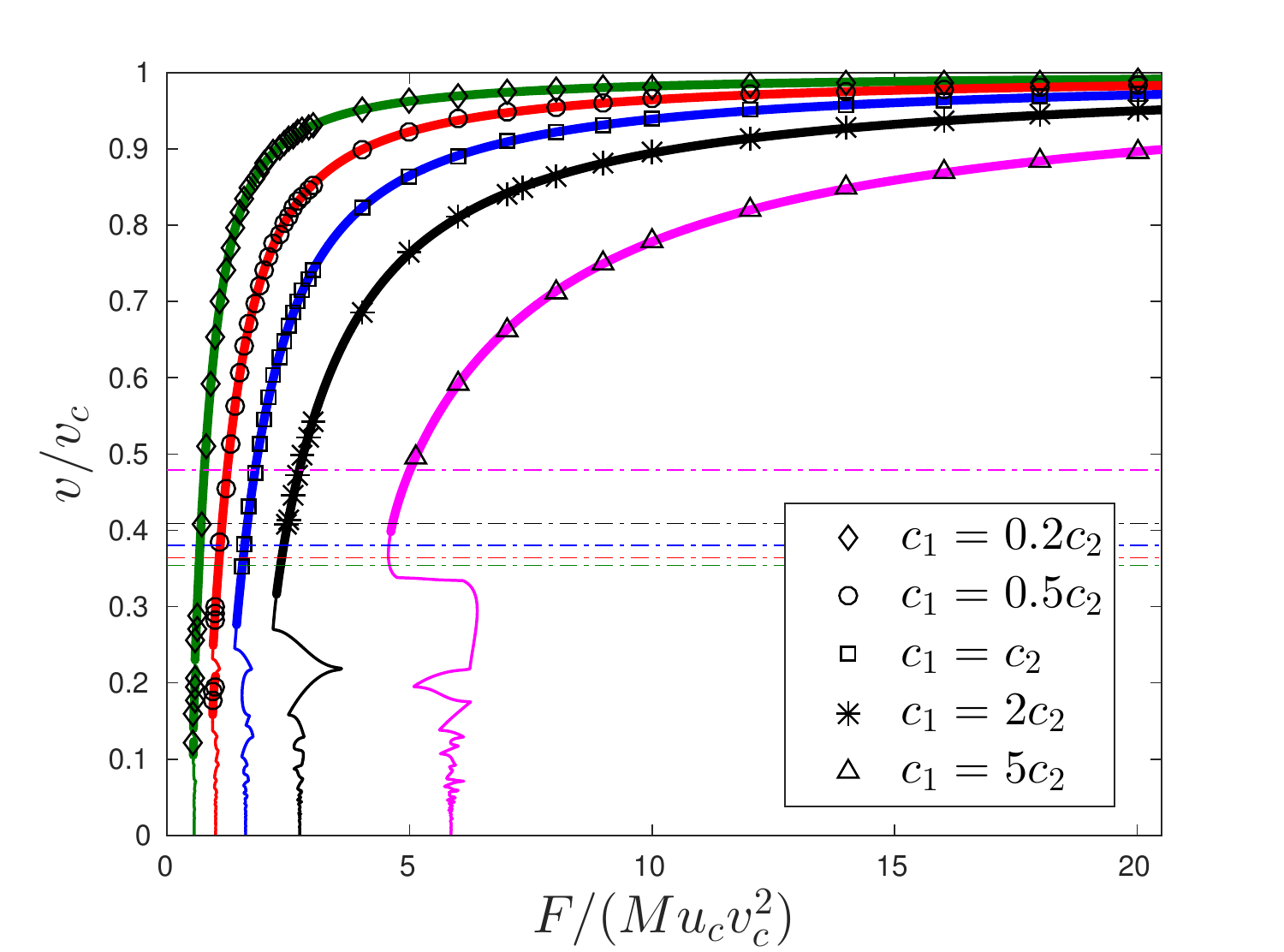} \\ a)}
\endminipage
\hfill
\minipage{0.5\textwidth}
\center{\includegraphics[width=\linewidth]{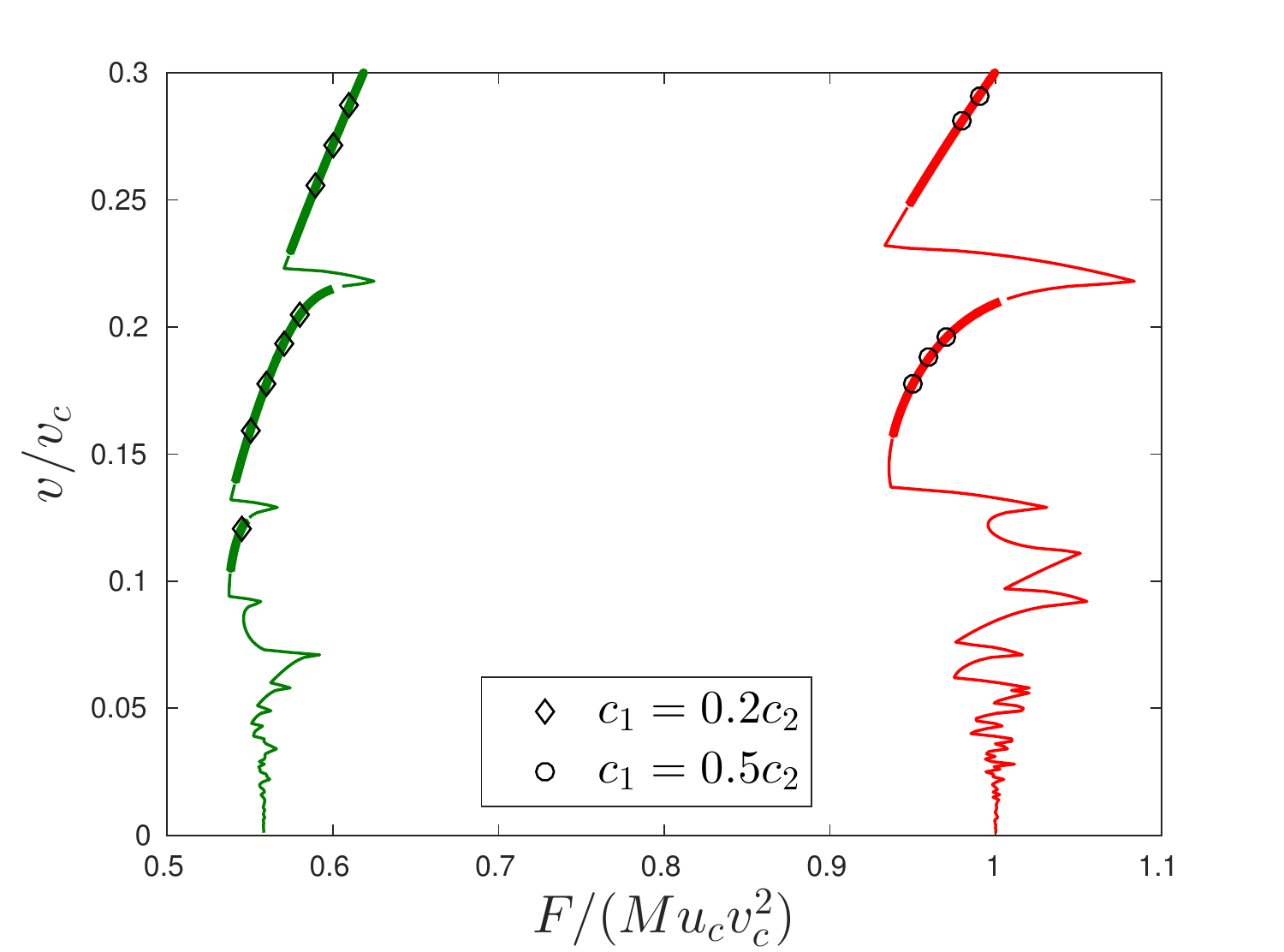} \\ b)}
\endminipage \\
\center{
\minipage{0.5\textwidth}
\center{\includegraphics[width=\linewidth]{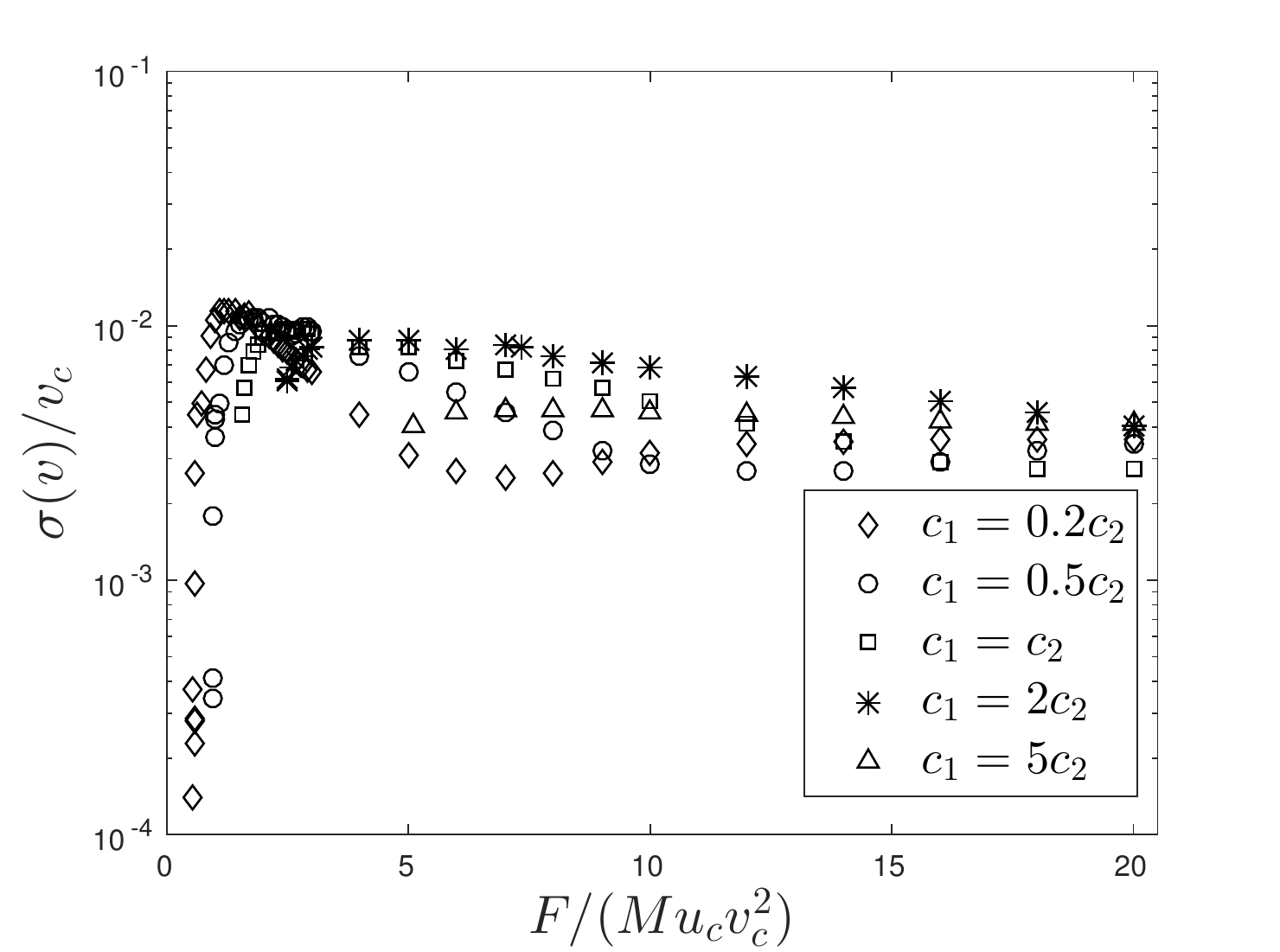} \\ c)}
\endminipage}
\captiondelim{. }
\caption[ ]{The results of the numerical simulations in the case of $v_f=0$ and several choices of $c_1$: a) Estimations of the steady-state crack speed, where the markers correspond to the calculated values from \eqref{eq:ComputedSpeed} whereas the solid lines are the calculations made by formula \eqref{eq:Dependence_F_vf_v}. The thick and thin lines correspond to the admissible and forbidden regimes, respectively. Dash-dotted lines correspond to the values $v_j^*$. b) The zoom of the plot in a) for the cases $c_1=0.2c_2,0.5c_2$, around the smaller values of $v$, c) Standard deviation from \eqref{eq:ComputedSpeed_2} of the estimates of the steady-state crack speed.}
\label{fig:Speed_Force_different_contrast}
\end{figure}

In this work we have also varied the spring stiffness $c_1$ describing the contraction properties of the horizontal and vertical springs. It has been highlighted in Section 3 that a change in the relative sizes of $c_1$ and $c_2$ led to qualitative changes in the admissible regimes. In Fig. \ref{fig:Speed_Force_different_contrast} we present the results of the evaluation of the steady-state crack speed, $v$, from the respective numerical simulations as compared with the corresponding analytical data.
Here we use the loading-speed relationship instead of the energy-speed diagram, as the former is characterised by a one-to-one relationship in most of the cases considered.

The simulations show that in the case of a weak interface it is always possible to reach a steady-state regime with a speed that is less than that which corresponds to the maximiser of the energy-speed diagram (i.e. $v<v_j^*$ from the Table \ref{table:TableMaximaERR}). Those values are indicated by dashed lines of the corresponding color on the relevant diagrams in Fig. \ref{fig:Speed_Force_different_contrast}a.

However, for two cases when the vertical links are much weaker than the horisontal ones ($c_1=0.2c_2$ and $c_1=0.5c_2$) there exist intervals in the admissible regimes that do not reflect the uniqueness in determination of crack speed. The last can be clearly seen in \ref{fig:Speed_Force_different_contrast}b). The numerical results presented on this figure were achieved using parameter set $\mathbb{S}_3$ from Table \ref{table:TableSettings}. With a high level of confidence, the simulations showed
that the solutions develop steady-state propagating regimes from few possible predicted admissible steady-state regimes (compare Fig.\ref{fig:Speed_Force_different_contrast}c)). However, we have not been unable to identify any rule explaining which regime is preferred, and why.

\begin{figure}[h!]
\minipage{0.5\textwidth}
\center{\includegraphics[width=\linewidth] {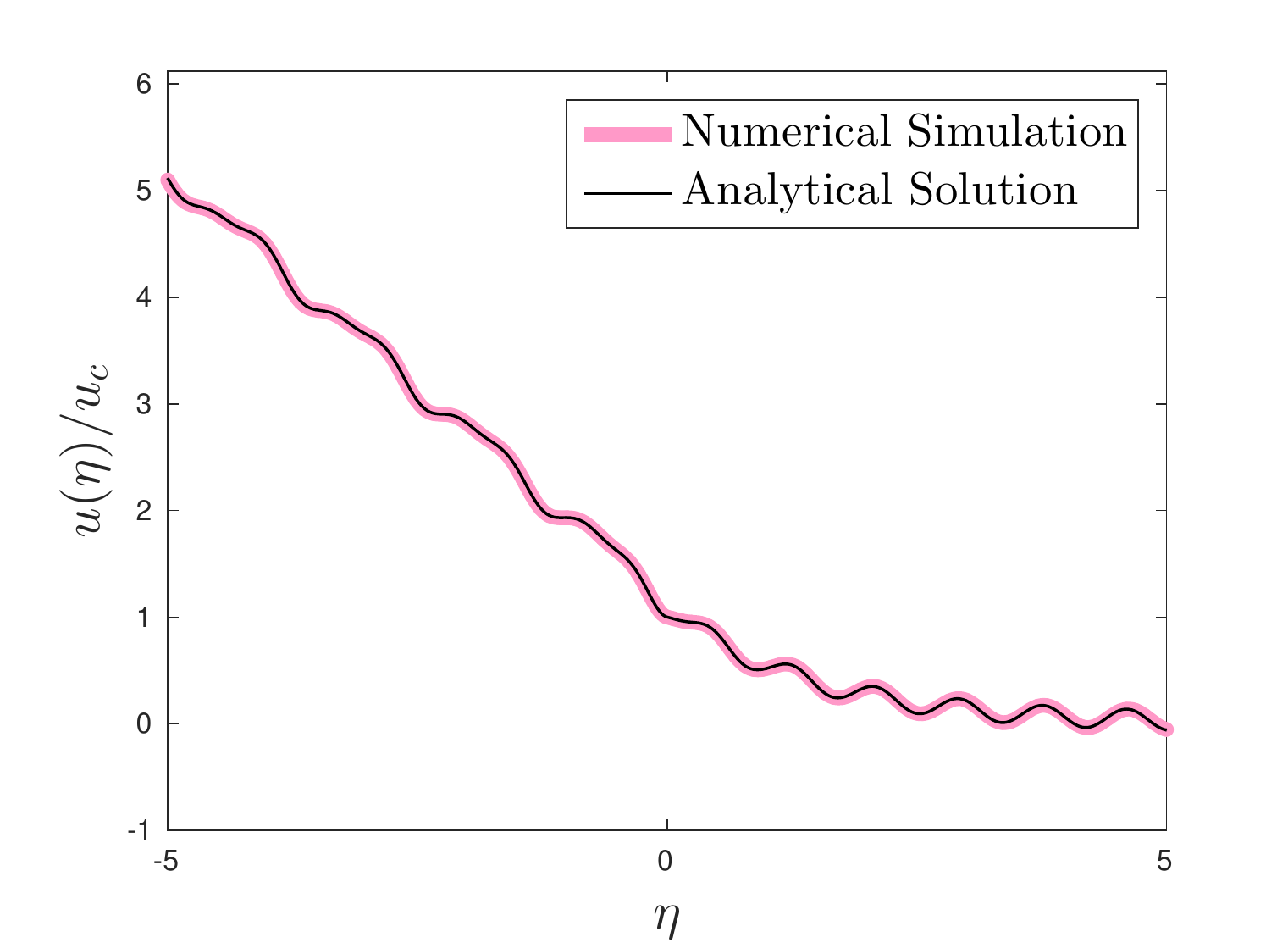} \\ a)}
\endminipage
\hfill
\minipage{0.5\textwidth}
\center{\includegraphics[width=\linewidth]{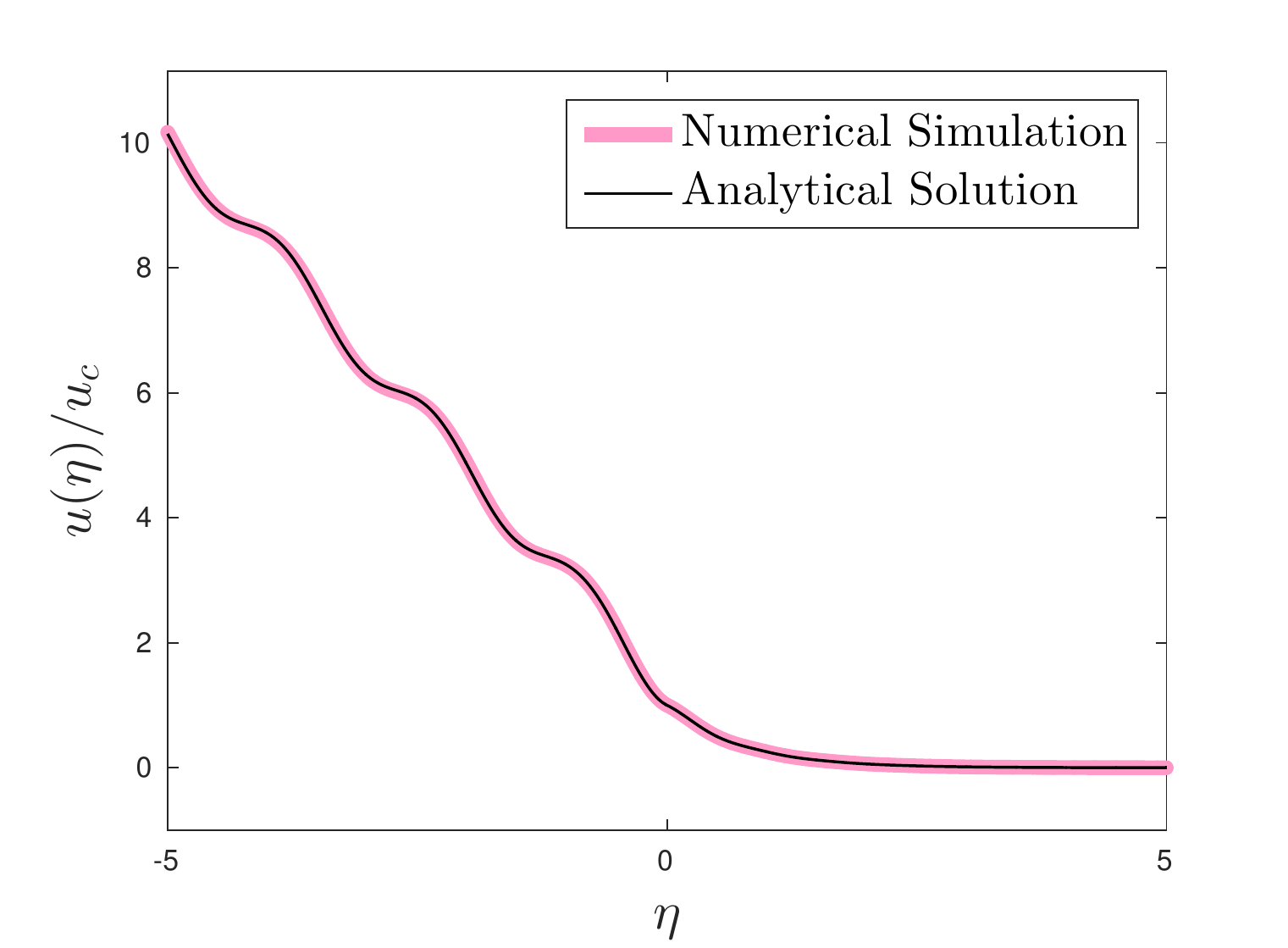} \\ b)}
\endminipage
\captiondelim{. }
\caption[ ]{The comparison of displacement fields for a fixed load ($v_f=0$): a) $c_1=0.5c_2$, $F=0.95Mu_cv_c^2$, $v=0.178v_c$, b) $c_1=2c_2$, $F=2.5Mu_cv_c^2$, $v=0.414v_c$.}
\label{fig:Displacements_Comparison}
\end{figure}

Finally, we point out some particular examples of the displacement profiles from the numerical simulations and compare them with their analytical equivalents.
These are shown in Fig.\ref{fig:Displacements_Comparison} and were chosen to illustrate the features of the radiating waves from the moving crack tip.
In the case where $c_1=0.5c_2$, $F=0.95Mu_cv_c^2$, we can observe the waves appearing behind and ahead of the crack tip, while in the second case $c_1=2c_2$, $F=2.5Mu_cv_c^2$, only waves behind the crack tip were initiated. Both computations were made with a fixed force position ($v_f=0$).

\section{Conclusions and Open Questions}

In the present paper we have provided an analysis of crack propagation and the development of a steady-state crack regime
in a double-chain structure. A Mode III fracture was initiated in the structure, and progressed by the application of a moving force initially situated far away from the crack tip.
A full analytical solution was derived, and the relationships between the steady-state crack speed and the loading parameters (force amplitude, $F$, and the speed, $v_f$, of the force movement) were evaluated. Moreover, accurate analysis of the analytical solution allowed us to separate the physically admissible and forbidden regimes of the steady-state fracture process (according to the fracture criterion \eqref{eq:FractureCondition_eta}).

The analytical results were supported by numerical simulations of the problem. We compared the results of these theoretical and numerical approaches and found excellent correlations in the examined cases. We showed that varying the numerical configurations and load implementations may affect the convergence rate of the solution to the steady-state regime. Although the instantaneous crack speed, $v(t_*)$, may exhibit different behaviour and depends on the limiting regime, the magnitude of the
steady-state crack speed, $v$, numerically defined as the mean value of the instantaneous speed, gives results equal to those derived analytically within the accuracy of both computations.

We especially point out the following results from the analysis:
\begin{itemize}
\item
A set of methods for initiation of an initiation speed, $v$, were identified analytically and confirmed numerically.
The convergence of the transient regime to its steady-state equivalent is insensitive to the particular choice of pair $(F,v_f)$ except when the speed $v_f$ is very small and non-zero.
The governing parameter is $F_*=F/(v_c-v_f)$ where $v_c$ is the speed of the waves propagating along the destroyed part of the structure. However, the questions of how and why the convergence to the limiting regime happens in various ways (see Fig.\ref{fig:DifferentNf}
Fig.\ref{fig:CrackSpeed_vs_Time} and Fig.\ref{fig:SameCrackSpeed}), and how it depends on the  steady-state, $v$, remain open.
\item
This confirms that the main determiner of the limiting regime is the amount of energy, $G$, introduced into the system, but not the particular load implementation, if the position of the force is sufficiently far from the crack tip. On the other hand, the distance between the initial position of the force and the crack tip may change with time, and the regime may still be stable even if this distance becomes smaller.
\item
The relationship between the steady-state crack speed, $v$, and the loading parameter $F_*$ is more more useful in the analysis than the energy release rate - speed diagram. This is because of the monotonic character of the function $v=v(F_*)$ for the admissible crack speeds.
As a result, there is some difficulty in making a choice of load parameters that leads to the desired steady-state fracture regime.
\item
A large difference in the elastic constants of the vertical and horizontal springs may lead to the existence of admissible regimes corresponding to a very "slow" steady-state movement that is not possible in the corresponding structure with more similar spring strengths. However, reducing the stiffnesses of the springs subjected to fracture leads to the appearance of a non-monotonic behaviour in the crack speed as a function of the applied load. This, in turn, causes uncertainty as to which regime would be expected to be the steady-state successor of the transient regime. This stiffness also makes a qualitative difference to the number of distinct intervals of admissible crack speeds.

\end{itemize}

There are a few more issues to highlight in this conclusion. We have not managed to reach all the admissible steady-state regimes in our numerical simulations,
but that was not the goal of our research.
We simply shown that it is possible to cover a larger region of the energy-speed diagram than expected by enriching the choices in applying loads.

The obtained numerical simulations prove that the steady-state regimes predicted by theory can be realised computationally.

We could consider a more complex loading system defined not by two constant parameters but instead two functions ($F(t)$, $v_f(t)$) which vary during the transient stage of the fracture process.
It is clear that the number of routes to any particular point on the energy-speed diagram, corresponding to a steady-state regime, by varying the loading configuration would then be infinite.

We could also whether a theoretically found admissible steady-state, however it is reached from the transient regime, is stable by virtue of a specific load. In this case, we might with start with initial conditions corresponding to that theoretically predicted steady-state. The efficiency of this strategy was proven in \cite{ayzenberg2014}, where it was used to analyse the spontaneous
steady-state propagation of a crack driven by internally accumulated energy.

The work in this article suggests that it might be useful to track fracture history before the steady-state regime is achieved. This might clarify some of the problems in obtaining the crack speeds at the borders of the admissible intervals.

In contrast with the results from \cite{marder1994}, no fracture events occur in the horizontal links ahead of the crack tip. The reason for this is rather simple and lies in the shape of the structure and the manner in which the load is applied. In our case, the moving force is situated far behind the crack tip while in \cite{marder1994} the structure was loaded along the entire boundary.

\vspace{2mm}

{\bf Acknowledgements} NG acknowledges support from the FP7 EU project CERMAT-2 (PITN-GA-2013-606878) and 
thanks to Dr M. Gromada for fruitful discussions during his secondment to CEREL. GM acknowledges partial support from the FP7 EU project  
TAMER (IRSES-GA-2013-610547) and from the Royal Society via Wolfson Research Merit Award.

\clearpage
\bibliographystyle{abbrv}
\bibliography{Bibliography}

\clearpage
\begin{center}
\bf\Large Supplementary material
\end{center}

\appendix
\section{Some technical details supporting evaluation of the solution}
\label{Appendix:Details}
\renewcommand{\theequation}{{\small SM}\arabic{equation}}

The material presented below is provided to aid understanding of the main technical steps used in derivation of the limit in the formula (40), which allows us to find the steady-state limit from the transient regime.

\subsection{Dispersion relations and the kernel function of the problem}
\label{Appendix:DispesionRelations}

\begin{figure}[h!]
\begin{minipage}[h]{0.5\linewidth}
\center{\includegraphics[width=1\linewidth]{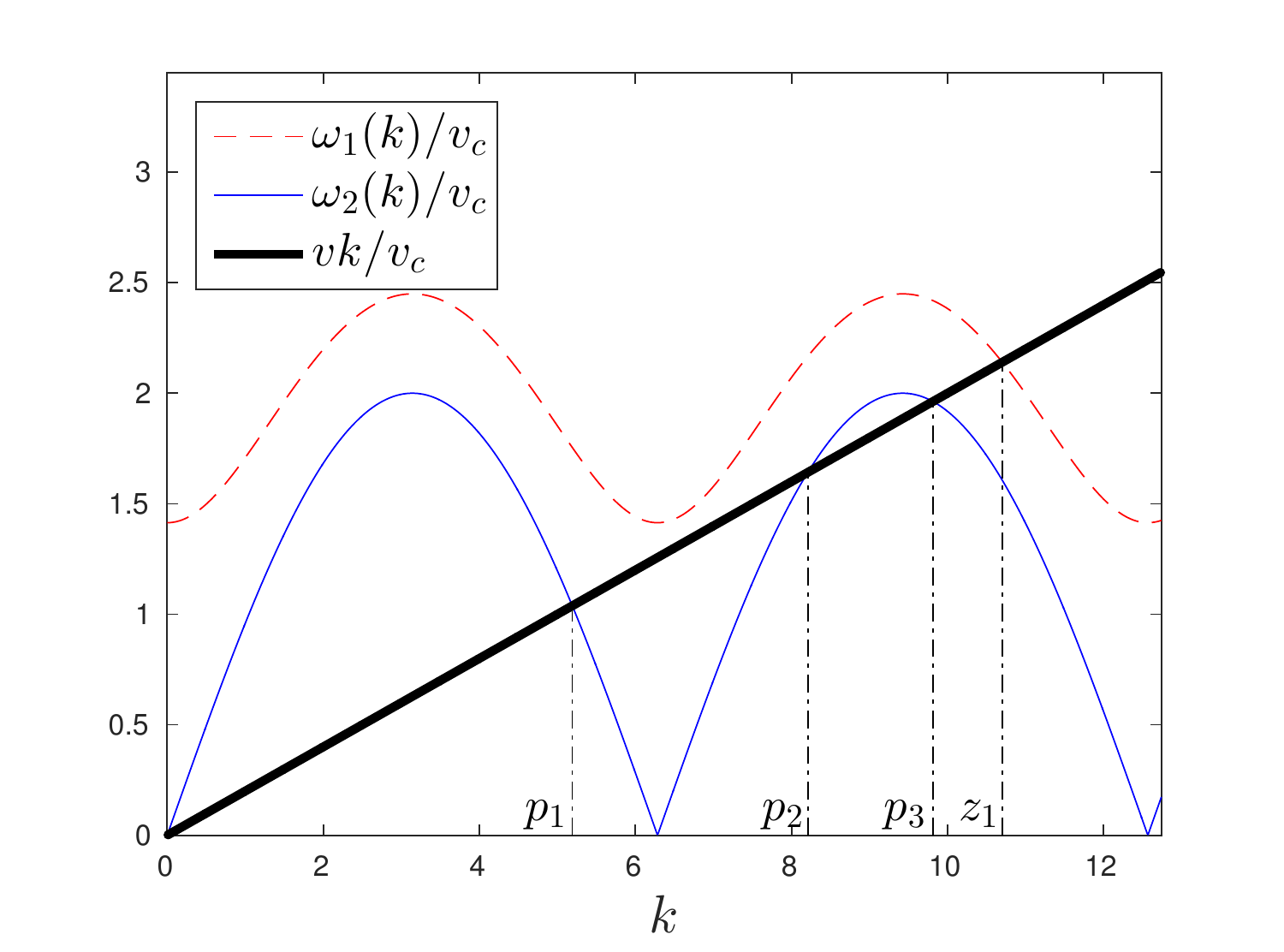} \\ a)}
\end{minipage}
\hfill
\begin{minipage}[h]{0.5\linewidth}
\center{\includegraphics[width=1\linewidth]{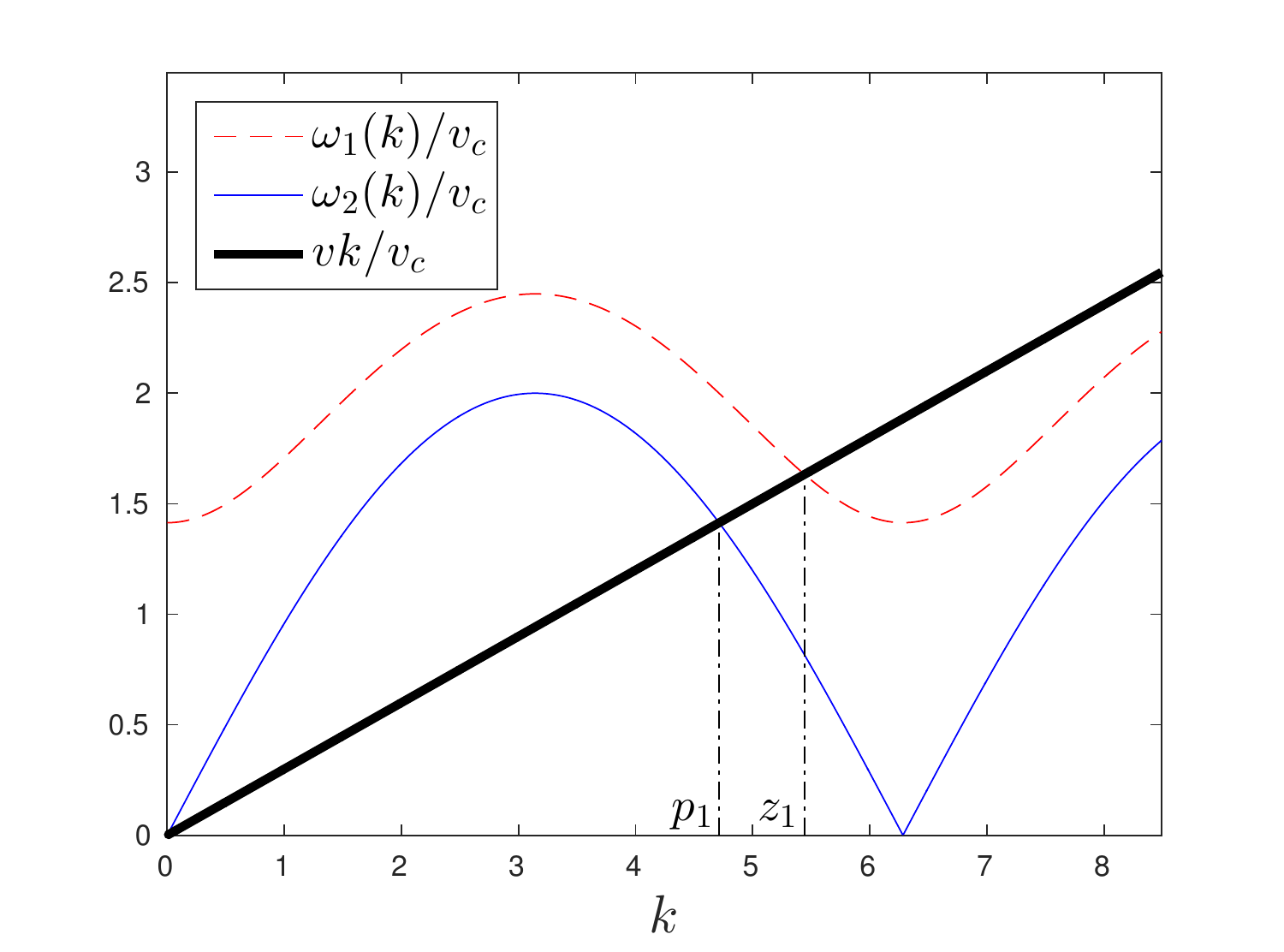} \\ b)}
\end{minipage}
\captiondelim{. }
\caption[ ]{ Dispersion diagram of a chain for several crack speeds, where $c_1=2c_2$: a) $v=0.2v_c$, b) $v=0.3v_c$. The dispersion relationships $\omega_1(k)$ and $\omega_2(k)$ (see (17)) correspond to an intact region of a chain and a broken one, respectively. The critical crack speed is $v_c=\sqrt{c_2/M}$.}
\label{fig:DispersionDiagram_Chain}
\end{figure}

The kernel function of the problem $L(k,s)$ in (20), and its steady-state limit $L(k)$ in (29), are given by:
\begin{equation}
L(k,s)=\frac{(s+ikv)^2+\omega_1^2(k)}{(s+ikv)^2+\omega_2^2(k)},\quad
L(k)=\lim_{s\to0+}L(k,s)=\frac{(0+ikv)^2+\omega_1^2(k)}{(0+ikv)^2+\omega_2^2(k)}
\label{L_functions}
\end{equation}
The objective here is to analyse the function $L(k)$ and its singularities, bearing in mind the properties of the function $L(k,s)$. From the definition of $L(k)$ it is clear that its real zeros and poles are intersection points of $\omega_1^2(k)$ and $\omega_2^2(k)$, as seen in (17), with the function $(vk)^2$, respectively. The plots of the dispersion relationships are presented in Fig.\ref{fig:DispersionDiagram_Chain} for several crack speeds.
Since $\omega_{1,2}^2(k)$ and $(vk)^2$ are even functions of $k$, it is sufficient to search for the positive roots of the equations $\omega_{1,2}(k)-vk=0$.
For the a fixed crack speed $v$ let us define:
\begin{equation}
\begin{gathered}
z_j>0:\quad \omega_1(q_j)-vz_j=0,\quad j=1, \ldots,Z,\\
p_j>0:\quad \omega_2(p_j)-vp_j=0,\quad j=1,\ldots,P,
\end{gathered}
\label{eq:ZerosPoles_Chain}
\end{equation}
where $P$ and $Z$ are integers which represent the total numbers of positive zeros and poles of function $L(k)$, respectively. There is also a root at the point $k=0$ in the function $\omega_2(k)$:
\begin{equation}
\omega_2(0)=0,
\end{equation}

By using the definition of $L(k,s)$, we can analyse the behaviour of the function $L(k)$ at its zeros and poles, as $k\to0$:
\begin{equation}
\begin{gathered}
(s+ikv)^2+\omega_1^2(k)\sim (v_c^2-v^2)(k-z^+(s))(k-z^-(s)),\quad k\to0,\\ (s+ikv)^2+\omega_2^2(k)\sim (v_c^2-v^2)(k-p^+(s))(k-p^-(s)),\quad k\to0,\\
\end{gathered}
\label{eq:LimitZero_DispersionRelations}
\end{equation}
where
\begin{equation}
z^{\pm}(s)=\frac{-isv\mp i\sqrt{\omega_0^2(v_c^2-v^2)+v_c^2s^2}}{v_c^2-v^2},\quad
p^{\pm}(s)=\frac{\mp i s}{v_c\mp v}.
\label{eq:ZerosPoles_at_zero}
\end{equation}
We can also find that:
\begin{equation}
z^{\pm}(s)=\frac{\mp i\omega_0}{\sqrt{v_c^2-v^2}}+O(s),\quad s\to 0.
\label{eq:ZerosPoles_at_zero_s0_a}
\end{equation}
Thus, function $L(k)$ has the zeros and poles defined in \eqref{eq:ZerosPoles_Chain}. The asymptotic relations hold (with the terms $s(k-z_i)$ or $s(k-p_i)$ omitted due to the limit $s\to0$):
\begin{equation}
\begin{gathered}
(s+ikv)^2+\omega_1^2(k)=(2z_iv-is)((\omega_1'(z_i)-v)(k-z_i)+is)+O((k-z_i)^2),\quad k\to z_i,\\
(s+ikv)^2+\omega_2^2(k)=(2p_iv-is)((\omega_2'(p_i)-v)(k-p_i)+is)+O((k-p_i)^2),\quad k\to p_i
\end{gathered}
\label{eq:AsymptoticsDispersion}
\end{equation}
Taking into an account that the functions $L^{\pm}(k)$ in \eqref{L_functions} should be free of zeros and poles in $\pm\Im{k}>0$, respectively,
we conclude that the function $L^+(k,s)$ contains zeros and poles of even index, i.e. $z_{2j},p_{2j},j=1,2,...$, whereas the function $L^-(k,s)$ contains zeroes and poles of odd index, i.e. $z_{2j-1},p_{2j-1},j=1,2,...$.

The asymptotic relationships for the functions $L^{\pm}(k,s)$ follow from \eqref{eq:ZerosPoles_at_zero}:
\begin{equation}
\begin{gathered}
L^+(k,s)=R\frac{k-z^+(s)}{k-p^+(s)},\quad L^-(k,s)=\frac{1}{R}\frac{k-z^-(s)}{k-p^-(s)},\quad k\to0,\quad s\to0,
\label{eq:AsymptoticsL+-AtZero}
\end{gathered}
\end{equation}
We write the asymptotic behaviour of $L^-(k,s)$ at positive zeros and poles as:
\begin{equation}
\begin{gathered}
L^-(k,s)=V_j^-((2z_{2j-1}v-is)((\omega_1'(z_{2j-1})-v)(k-z_{2j-1})+is))+O((k-z_{2j-1})^2),\quad k\to z_{2j-1}\\
L^-(k,s)=\frac{W_j^-}{(2p_{2j-1}v-is)((\omega_1'(z_{pj-1})-v)(k-p_{2j-1})+is)}+O((k-p_{2j-1})^2),\quad k\to p_{2j-1},
\end{gathered}
\label{eq:AsymptoticsL-AtPositiveRoots}
\end{equation}
where $V_j^-,W_j^-$ are some constants.

\subsection{Analysis of the right-hand side of equation (19)}
\label{Appendix:ModificationOfRightPart}

Multiplying one of the functions on the right-hand side of equation (19) by $s$ yields:
\begin{equation}
\frac{F}{M}\frac{s}{s+ik(v-v_f)}\frac{e^{-ikn_0}}{[(s+ikv)^2+\omega_2^2(k)]L^-(k,s)}
\label{eq:RightPart}
\end{equation}
The key point to notice here is that the only non-zero values of this function in the limit $s\to0+$, are associated with the zeros of the denominator of the last fraction:

\begin{equation}
\left[(s+ikv)^2+\omega_2^2(k)\right]L^-(k.s)=0,\quad k=z^-(s),p^+(s),z_{2j-1},p_{2j},\,j=1,2,..., \quad s\to0.
\label{eq:FunctionOfRightPart}
\end{equation}
Now, let us consider the additional limit $k\to0$:
\begin{equation}
\frac{e^{-ikn_0}}{\left[(s+ikv)^2+\omega_2^2(k)\right]L^-(k,s)}=-\frac{R}{v_c^2-v^2}\frac{1}{z^-(s)}\frac{1}{k-p^+(s)}+O(1),\quad k\to0,\,s\to0.
\label{eq:formula_aux_1}
\end{equation}
The product of the middle fraction of \eqref{eq:RightPart} and the term $1/(k-p^+(s))$ gives:
\begin{equation}
\frac{s}{s+ik(v-v_f)}\frac{1}{k-p^+(s)}=\frac{1}{i(v-v_f)}\frac{s}{p^+(s)-\frac{is}{(v-v_f)}}\left[\frac{1}{k-p^+(s)}-\frac{1}{k-\frac{is}{(v-v_f)}}\right].
\label{eq:representation_0}
\end{equation}
We can then use expression for $p^+(s)$ from \eqref{eq:ZerosPoles_at_zero} to obtain:
\begin{equation}
\frac{1}{i(v-v_f)}\frac{s}{p^+(s)-\frac{is}{(v-v_f)}}=\frac{v_c-v}{v_c-v_f}
\end{equation}
The last expression does not depend on $s$ and is only zero when $v=v_c$. We finally obtain:
\begin{equation}
\frac{s}{s+ik(v-v_f)}\frac{1}{k-p^+(s)}=
\frac{v_c-v}{v_c-v_f}\left[\frac{1}{k+i0}-\frac{1}{k-i0}\right]+ O(s),\quad s\to0+.
\label{eq:representation_1}
\end{equation}
This expression is valid for $v_f<v$ but as long as the location of the force remains behind the crack tip this suitable for our purposes.

We shall now consider the limits $k\to z_{{2j-1}}$, as mentioned in \eqref{eq:FunctionOfRightPart}:
\begin{equation}
\begin{gathered}
\frac{e^{-ikn_0}}{\left[(s+ikv)^2+\omega_2^2(k)\right]L^-(k)}=
\frac{\hat{V}_j^-e^{-in_0z_{2j-1}}}{(\omega_1'(z_{2j-1})-v)(k-z_{2j-1})+is}
+O(1),\quad k\to z_{2j-1},\\
\hat{V}_j^-=\frac{1}{\left[(s+iz_{2j-1}v)^2+\omega_2^2(z_{2j-1})\right]}
\frac{1}{V_j^-(2z_{2j-1}v-is)}=O(s),\quad s\to 0
\end{gathered}
\end{equation}
We note that the product of the middle fraction of \eqref{eq:RightPart} and the factor in the last expression is:
\begin{equation}
\begin{gathered}
\frac{s}{s+ik(v-v_f)}\frac{1}{(\omega_1'(z_{2j-1})-v)(k-z_{2j-1})+is}\\
=\frac{1}{i(v-v_f)(\omega_1'(z_{2j-1})-v)}\frac{s}{z_{2j-1}-\frac{is}{(\omega_1'(z_{2j-1})-v)}-\frac{is}{v-v_f}}\left[\frac{1}{k-z_{2j-1}+\frac{is}{(\omega_1'(z_{2j-1})-v)}}-\frac{1}{k-\frac{is}{v-v_f}}\right]
\end{gathered}
\end{equation}
In this case, the factor in front of the square brackets is:
\begin{equation}
\frac{s}{z_{2j-1}-\frac{is}{(\omega_1'(z_{2j-1})-v)}-\frac{is}{v}}=o(s),\quad s\to0.
\end{equation}
In other words, there is no contribution from the points $k=z_{2j-1}$ to the expression \eqref{eq:RightPart} in the limit $s\to0$. The same reasoning applies to the limit $k\to p_{2j}$ which are also roots of function in \eqref{eq:FunctionOfRightPart}.

It should be stressed that in the case $v=v_f$ the factorisation should be accomplished differently.
Indeed, in this case the fraction $s/(s+ik(v-v_f))=1$, which leads to certain changes in the analysis, e.g. in \eqref{eq:representation_1}.

We can observe that function the $H_0(k,s)$ in (19) comes from the initial conditions of the original problem, does not contain any singularities, and that $sH_0(k,s)=o(s), s\to0$. Thus, using the expressions from \eqref{eq:ZerosPoles_at_zero} and \eqref{eq:ZerosPoles_at_zero_s0_a} and the reasoning associated with the above function \eqref{eq:RightPart} we conclude that the function on the right-hand side of equation (19) weakly converges:
\begin{equation}
\begin{gathered}
\frac{Fe^{-ikl_0}}{M}\frac{s}{s+ik(v-v_f)}\frac{1}{(s+ikv)^2+\omega_2^2(k)}\frac{1}{L^-(k,s)}+\frac{sH_0(k,s)}{(s+ikv)^2+\omega_2^2(k)}\frac{1}{L^-(k,s)}\\
= \frac{C}{0-ik}+\frac{C}{0+ik} +O(s),\quad s\to0,\\
C=\frac{F}{M}\frac{v_c-v}{v_c-v_f}\frac{R}{\sqrt{\omega_0^2(v_c^2-v^2)}},
\end{gathered}
\label{eq:Final_RightPart}
\end{equation}
where the quantity $\omega_0^2$ is defined in (17).

\end{document}